%% file: main.tex
\documentclass[letterpaper, 10 pt, onecolumn, conference]{ieeeconf}  
\IEEEoverridecommandlockouts                              
\input{def}

\title{The Value of Compromising Strategic Intent in General Lotto Games}

\author{
    Gilberto D\'iaz-Garc\'ia$^{*}$,
    Keith Paarporn$^{**}$,
    Jason R. Marden$^{*}$
    \thanks{$^{*}$Department of Electrical \& Computer Engineering, University of California, Santa Barbara, Santa Barbara, CA (e-mail: gdiaz-garcia@ucsb.edu, jrmarden@ucsb.edu ) }
    \thanks{$^{**}$Department of Computer Science, University of Colorado, Colorado Springs, CO (e-mail: kpaarpor@uccs.edu) }
}

\begin{document}

\maketitle

\begin{abstract} 
    Resource allocation in adversarial environments is a fundamental challenge across various domains, from corporate competition to military strategy. This article examines the impact of compromising an opponent's strategic intent in the context of General Lotto games, a class of resource allocation problems. We consider a scenario where one player, termed the ``Breaker," has access to partial information about their opponent's strategy through a binary sensor. This sensor reveals whether the opponent's allocated resources exceed a certain threshold. Our analysis provides a comprehensive characterization of equilibrium strategies and payoffs for both players under this information structure. Through numerical studies, we demonstrate that the information provided by the sensor can significantly improve the Breaker's performance. 
\end{abstract}

\IEEEpeerreviewmaketitle

%%%%%%%%%%%%%%%%%%%%%%%%%%%%%%%%%%%%%%%%%%%%%%%%%%%%%%%%%%%%%%%%%%%%%%%%%%%%%%%%%%%%%%%%%%%%%%%%%%%%%%%%%%%%%%%%%%%%%%%%%%%%%%%%%%%%%%%%%%%%%%%%%%%%%%%%%%%%%%%%%%%%%%%%%%%%%%%%%%%%%%%%%%%%%%%%%%%%%%%%%%%%%%%%%%%%%%%%%%%%%%%%%%%%%%%%%%%%%%%%%%%%%%%%%%%%%%%%%
\section{Introduction}

Resource allocation in competitive environments is a fundamental challenge that shapes the outcomes of adversarial interactions across various domains, from corporate rivalries to military strategy. Entities in these environments continuously seek opportunities to gain a competitive edge, leveraging various avenues such as superior information, strategic unpredictability, or tactical deception. This paper focuses on one critical aspect of competitive advantage: the ability to unveil elements of an opponent's strategic intent. The significance of such information has been demonstrated throughout history, from the Allied forces breaking the Enigma code in World War II to the execution of Mary, Queen of Scots after the discovery of her encrypted assassination plans.

This paper explores the impact of compromising strategic intent in a well-studied class of competitive resource allocation problems known as General Lotto games \cite{}. These games, a popular variant of the Colonel Blotto model \cite{}, provide a rich framework for analyzing strategic environments involving two players competing over a given contest. Researchers have extensively analyzed General Lotto games, establishing equilibrium characterizations for various scenarios \cite{}, including those with incomplete and asymmetric information \cite{}, networked contest environments \cite{}, and multi-player $(>2)$ settings \cite{}. Of particular interest has been the study of General Lotto games with asymmetric information. This line of research extends the classical framework to scenarios where players possess different levels of information about contest parameters or opponent capabilities, significantly altering the strategic landscape and equilibrium outcomes \cite{}. Interestingly, these works demonstrate that informational superiority can significantly advance a player's strategic position.

\begin{figure}[htb]
    \centering
    \includegraphics[width=0.45\textwidth]{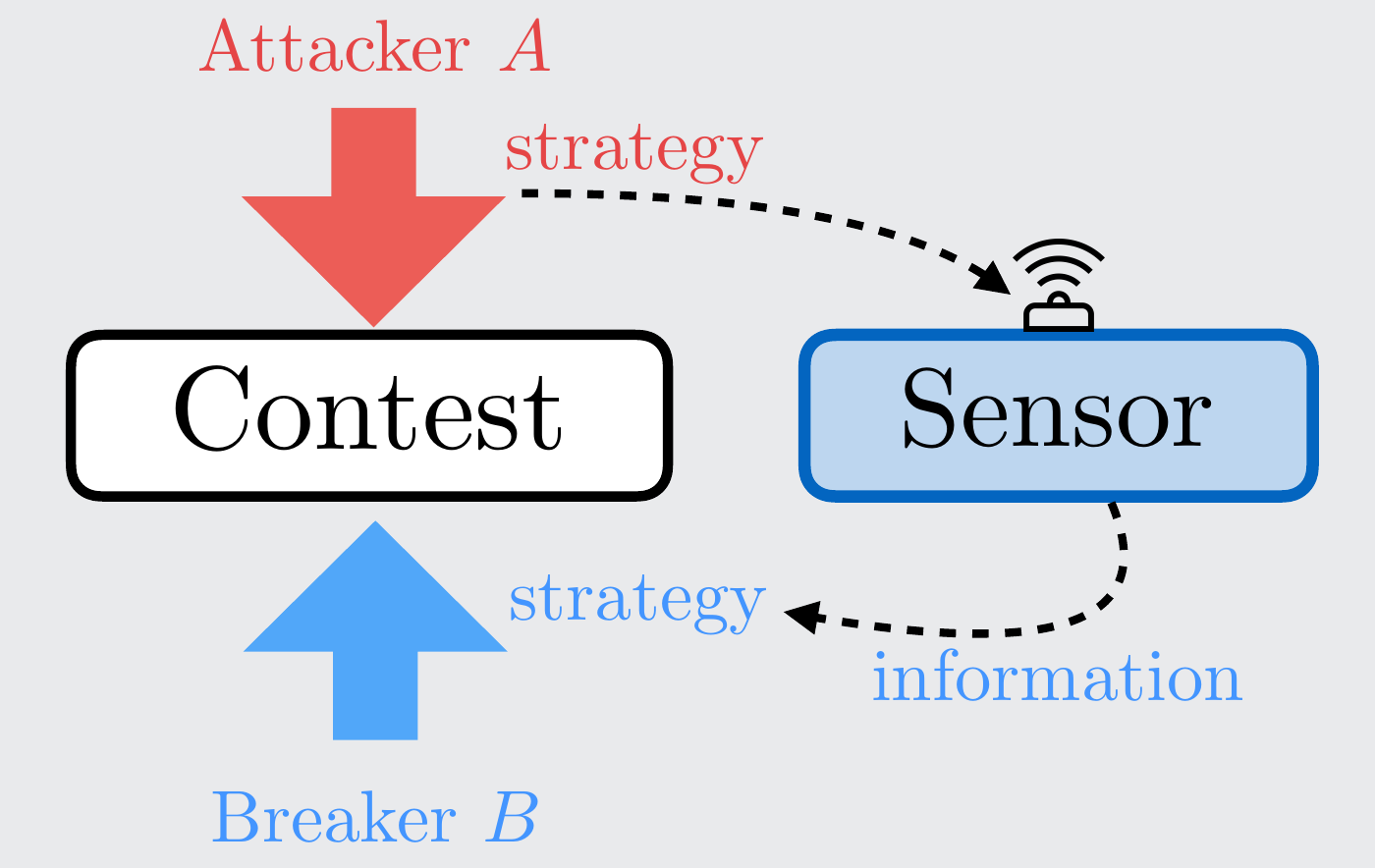}
    \caption{The highlighted scenario captures a strategic environment where two players, referred to as the Attacker and Breaker, must decide how to allocate their limited assets ($X_A$ and $X_B$ respectively) over a given contest. In the classic General Lotto setting, each agent simultaneously chooses their allocation strategy to optimize their performance guarantees without any knowledge of their opponent's allocation strategy. This work extends this classic framework by introducing an information asymmetry that favors the Breaker as highlighted above. Specifically, we consider a setting where the Breaker is able to observe a binary signal indicating whether the Attacker's allocated assets, i.e., the realization of the Attacker's randomized allocation strategy, are above or below a given threshold value. The Breaker can then condition their allocation strategy on this observed information.}
    \label{fig:diagram}
\end{figure}

Our work contributes to this growing body of literature by examining a specific form of information asymmetry: one player's ability to partially unveil the strategic intent of their opponent. The specific scenario considered in this paper is depicted in Figure~\ref{fig:diagram}, where two players, referred to as the Attacker and Breaker, must decide how to allocate their limited assets, denoted as $X_A$ and $X_B$ respectively, over a given contest.

In the classic General Lotto setting, each agent simultaneously chooses their allocation strategy to optimize their performance guarantees. Typically, both players have full information about the problem parameters, such as the quantity of assets available to each player and the contest valuations. Our work extends this framework by introducing an information asymmetry that favors the Breaker. Specifically, we allow the Breaker to observe a binary signal indicating whether the Attacker's allocated assets are above or below a given threshold value. The Breaker can then condition their allocation strategy on this observed information. The goal of this study is to understand the

This paper investigates how this partial information about the Attacker's strategy impacts the resulting equilibria and performance guarantees. By analyzing this scenario, we aim to provide insights into the strategic value of compromising an opponent's intent in such competitive resource allocation problems. The main contribution of this paper is given in Theorem~\ref{thm:info}, where we provide a comprehensive analytical characterization of equilibrium strategies and payoffs for both players in our General Lotto game with a binary sensor. It is important to highlight that this characterization holds for any threshold value and any combination of player resource budgets $X_A$ and $X_B$. Second, in Section~\ref{sec:sims} we provide a series of numerical experiments that seek to quantify the performance improvements the Breaker achieves using the binary threshold sensor, compared to the scenario without such information. These results clearly demonstrate how compromising an opponent's strategic intent can significantly alter the competitive landscape. While our analysis centers on binary information, it lays the groundwork for investigating more complex information structures in competitive resource allocation problems.

%%%%%%%%%%%%%%%%%%%%%%%%%%%%%%%%%%%%%%%%%%%%%%%%%%%%%%%%%%%%%%%%%%%%%%%%%%%%%%%%%%%%%%%%%%%%%%%%%%%%%%%%%%%%%%%%%%%%%%%%%%%%%%%%%%%%%%%%%%%%%%%%%%%%%%%%%%%%%%%%%%%%%%%%%%%%%%%%%%%%%%%%%%%%%%%%%%%%%%%%%%%%%%%%%%%%%%%%%%%%%%%%%%%%%%%%%%%%%%%%%%%%%%%%%%%%%%%%%
\section{Model}\label{sec:model}

We first present a preliminary on the classic General Lotto game as introduced in \cite{Hart_2008}. We then extend this setup to incorporate an information-gathering mechanism  (Figure \ref{fig:diagram}) takes the form of a binary threshold sensor.

\subsection{The General Lotto Game}

Consider two players $i \in \crl{A,B}$, each one with a limited budget of resources $X_A,X_B>0$ respectively. They compete over a contest of unit value. To win, a player needs to allocate more resources to the contest than the opponent. An admissible strategy for player $i$ is any randomization over allocations $x_i \geq 0$, such that it does not exceed its budget $X_i$ \emph{in expectation}. Specifically, an admissible strategy for player $i$ is any cumulative distribution $F_i$ over $\R_{\geq 0}$ belonging to $\F\prth{X_i}$, defined as 
\begin{equation}\label{eq:lotto_constraint}
    F_i \in \F\prth{X_i} \iff \Ex_{ x_i \sim F_i }\sqr{ x_i } \leq X_i.
\end{equation}
Note that it is possible to select a strategy $F_i$ where an allocation $x_i$ drawn from $F_i$ violates the budget, i.e. $x_i > X_i$. However, $F_i \in \F\prth{X_i}$ requires the expectation over all allocations $x_i \geq 0$ does not exceed $X_i$.

Given a pair of admissible strategies $(F_A,F_B)$, the expected utility to player $B$ is defined to be
\begin{equation}\label{eq:GL_utility}
    U_B( F_A, F_B ) := \Ex_{ x_A \sim F_A, x_B \sim F_B } \sqr{ \Ind\crl{ x_A \leq x_B } }
\end{equation}
where $\Ind\crl{\cdot}$ is the indicator function that takes value 0 if false and 1 if true. Consequently, the expected utility to player $A$ is given by $U_A(F_A,F_B) := 1 - U_B(F_A,F_B)$.

Here, we are assuming that in case of a tie, player $B$ wins. The utility for each player represents the probability of winning the contest. This simultaneous-move game is called the \emph{General Lotto game}, and we denote it with $\GL(X_A,X_B)$.

% For simplicity and without losing any generality, we assume that the value of the contest is normalized, i.e., $v=1$.
% Therefore, the General Lotto game is a constant-sum game with value equal to $v$ and, in our particular case, equal to $1$. 

% We are interested in the resulting behavior that emerge from this setting. Particularly, in the equilibrium payoff that both players will obtain for any particular instance of a General Lotto game. 

We say a strategy profile $(F_A^*,F_B^*)$ is an \emph{equilibrium} if for any $F_A \in \F\prth{X_A}$ and $F_B \in \F\prth{X_B}$,
\begin{equation*} 
    U_A(F_A^*,F_B^*) \geq U_A(F_A,F_B^*) \text{ and } U_B(F_A^*,F_B^*) \geq U_A(F_A^*,F_B).
\end{equation*}

The equilibrium of $\GL(X_A,X_B)$ was first established in \cite{Hart_2008}. We summarize the result below, which provides the payoffs the players obtain in an equilibrium.

\begin{theorem}[\cite{Hart_2008}]\label{thm:GL}
    Consider the General Lotto game $\GL(X_A,X_B)$. Then in an equilibrium, player $B$ obtains payoff
    \begin{equation} \label{eq:lotto} \begin{aligned}
    U_B^*(X_A,X_B) &:= \begin{cases}
                        \frac{X_B}{2 X_A}, & \text{ if $ X_B < X_A$ } \\
                        1 - \frac{X_A}{2 X_B}, & \text{ if $ X_B \geq X_A$ }
                    \end{cases}
\end{aligned} \end{equation}
and player $A$  obtains payoff $1 - U_B^*(X_A,X_B)$.
\end{theorem}

Referring to Figure \ref{fig:diagram}, the game $\GL(X_A,X_B)$ represents only the contest portion of the diagram, i.e. without any information sensor. In the following, we now formulate an extension that incorporates an information-gathering component for player $B$ that signals information about player $A$'s allocation.

\subsection{General Lotto Games with Binary Information}

Player $B$ (the ``Breaker") receives a signal $s$ from a sensor about the \emph{realized allocation} $x_A \sim F_A$ from the strategy of player $A$ (the ``Attacker") before the contest takes place. The signal will reveal whether $x_A$ was above or below a fixed threshold value $\tau>0$. Thus, the signal takes a binary value: $s=0$ if $x_A \leq \tau$ and $s=1$ if $x_A > \tau$. The Breaker $B$ is then able to select a strategy $F_{B|0} \in \F\prth{X_B}$ or $F_{B|1} \in \F\prth{X_B}$ contingent on which signal was received. The sequence of events is summarized as follows.

\begin{enumerate}
    \item Player $A$ chooses its strategy $F_A \in \F\prth{X_A}$. 
    \item An allocation $x_A \sim F_A$ for player $A$ is realized. It generates signal $s=0$ if $x_A \leq \tau$, and signal $s=1$  if $x_A > \tau$.
    \item Player $B$ utilizes strategy $F_{B|s}$.
\end{enumerate}

Here, we are assuming that player $A$ selects $F_A$ with the knowledge of the value $\tau$. We also assume both players seek to optimize their expected payoffs in regards to this sequence. In particular, the expectation is taken with respect to the randomness of $F_A$ and $F_B = (F_{B|0},F_{B|1})$, yielding an expected payoff of
\begin{equation}\label{eq:GLI_utility}
    \begin{aligned}
        &U_B(F_A,F_B,\tau) := \\
        &\Ex_{ x_A \sim F_A} \left[ \Ind\crl{ x_A \leq \tau } \Ex_{x_B\sim F_{B|0}} \sqr{\Ind\crl{x_B \geq x_A}} \right. \\
        &\quad + \left. \Ind\crl{ x_A > \tau } \Ex_{x_B\sim F_{B|1}} \sqr{\Ind\crl{x_B \geq x_A}} \right]
    \end{aligned}
\end{equation}
for the Breaker $B$, and an expected payoff of $U_A(F_A,F_B,\tau) := 1 - U_B(F_A,F_B,\tau)$ for the Attacker $A$. This defines an instance of a \textit{General Lotto game with Binary Information}, which we denote as $\GLI(X_A,X_B,\tau)$. We are interested in deriving analytical characterizations of the equilibrium payoffs and strategies of $\GLI(X_A,X_B,\tau)$.

% In addition to the previously described scenario, we now allow to player $B$ to gather information about its adversary. This information is represented by a threshold value $\tau>0$. This means that player $B$ will know with total certainty if the realization of player's $A$ distribution is above or below such threshold. However, player $A$ is aware of the value of $\tau$ and can choose its strategy accordingly.  and we can summarize its steps as follows, 

% This will result in an equilibrium winning rule $\GLI(X_A,X_B,\tau)$. We devote the rest of this document to characterize the equilibrium of the aforementioned scenario and the function $\GLI(X_A,X_B,\tau)$.

%%%%%%%%%%%%%%%%%%%%%%%%%%%%%%%%%%%%%%%%%%%%%%%%%%%%%%%%%%%%%%%%%%%%%%%%%%%%%%%%%%%%%%%%%%%%%%%%%%%%%%%%%%%%%%%%%%%%%%%%%%%%%%%%%%%%%%%%%%%%%%%%%%%%%%%%%%%%%%%%%%%%%%%%%%%%%%%%%%%%%%%%%%%%%%%%%%%%%%%%%%%%%%%%%%%%%%%%%%%%%%%%%%%%%%%%%%%%%%%%%%%%%%%%%%%%%%%%%
\section{Results}\label{sec:result}

In this section, we present our main result which comprehensively characterizes the equilibrium payoffs of $\GLI(X_A,X_B,\tau)$ for all possible parameter tuples $(X_A,X_B,\tau)$.

\begin{theorem} \label{thm:info}
    For an instance of the General Lotto game with Binary Information $\GLI(X_A,X_B,\tau)$ the player $B$'s equilibrium payoff $U_B^*(X_A,X_B,\tau)$ is given as follows:    
    \begin{enumerate}
        \item Suppose $X_B \leq \frac{\tau}{2}$. Then, the equilibrium payoff is given by,
        \begin{align}
            &1 - \frac{X_A}{2 X_B},                                     && \text{ if } X_A \leq X_B                                \tag{I}    \label{eq:ua:i} \\
            &\frac{X_B}{2X_A},                                          && \text{ if } X_B \leq X_A \leq \frac\tau2                \tag{II}   \label{eq:ua:ii} \\
            &\frac{X_B}{\tau}\prth{ 1 - \frac{2X_A-\tau}{3\tau} },      && \text{ if } \frac\tau2 \leq X_A \leq \frac{5\tau}{4}    \tag{III}  \label{eq:ua:iii} \\
            &\frac{X_B}{X_A + \sqrt{ X_A^2 - \tau^2 }},                 && \text{ if } \frac{5\tau}{4} \leq X_A                    \tag{IV}   \label{eq:ua:iv}
        \end{align} 

        \item Suppose $\frac\tau2 \leq X_B \leq \frac{3\tau}{4}$. Then, the equilibrium payoff is given by,
        \begin{align}
            &1 - \frac{2X_A}{\tau}\prth{ 1 - \frac{X_B}{\tau} },                        && \text{ if } X_A \leq \frac\tau2                        \tag{V}   \label{eq:ua:v} \\
            &\frac{2X_B \prth{ 2\tau - X_A }}{3\tau^2},                                 && \text{ if } \frac\tau2 \leq X_A \leq \frac{5\tau}{4}   \tag{VI}  \label{eq:ua:vi} \\ 
            &\frac{X_B}{X_A + \sqrt{ X_A^2 - \tau^2 }},                                 && \text{ if } \frac{5\tau}{4} \leq X_A                   \tag{VII} \label{eq:ua:vii}
        \end{align} 
        
        \item Suppose $\frac{3\tau}{4} \leq X_B$. Then, the equilibrium payoff is given by,
        \begin{align}
            &1 - \frac{X_A}{X_B + \sqrt{X_B^2 + \tau^2}},       && \text{ if } X_A \leq \sqrt{ X_B^2 + \tau^2 } \tag{VIII}   \label{eq:ua:viii} \\
            &\frac{X_B}{X_A + \sqrt{ X_A^2 - \tau^2 }},         && \text{ if } X_A \geq \sqrt{ X_B^2 + \tau^2 } \tag{IX}     \label{eq:ua:ix}
        \end{align} 
    \end{enumerate}
\end{theorem}

While the statement of Theorem~\ref{thm:info} above provides only the equilibrium payoffs, the corresponding equilibrium strategies $(F_A^*,F_B^*)$ are derived via the proofs found in the Appendix. We omit their characterizations in the Theorem statement for clearer and more concise exposition. We elaborate on our proof approach in Section~\ref{sec:analysis}. 

\section{Numerical Simulations}\label{sec:sims}
\begin{figure}[hbt]
    \centering
    \subfloat[ General Lotto game \label{fig:nex:gl}                            ]{\includegraphics[height=20ex]{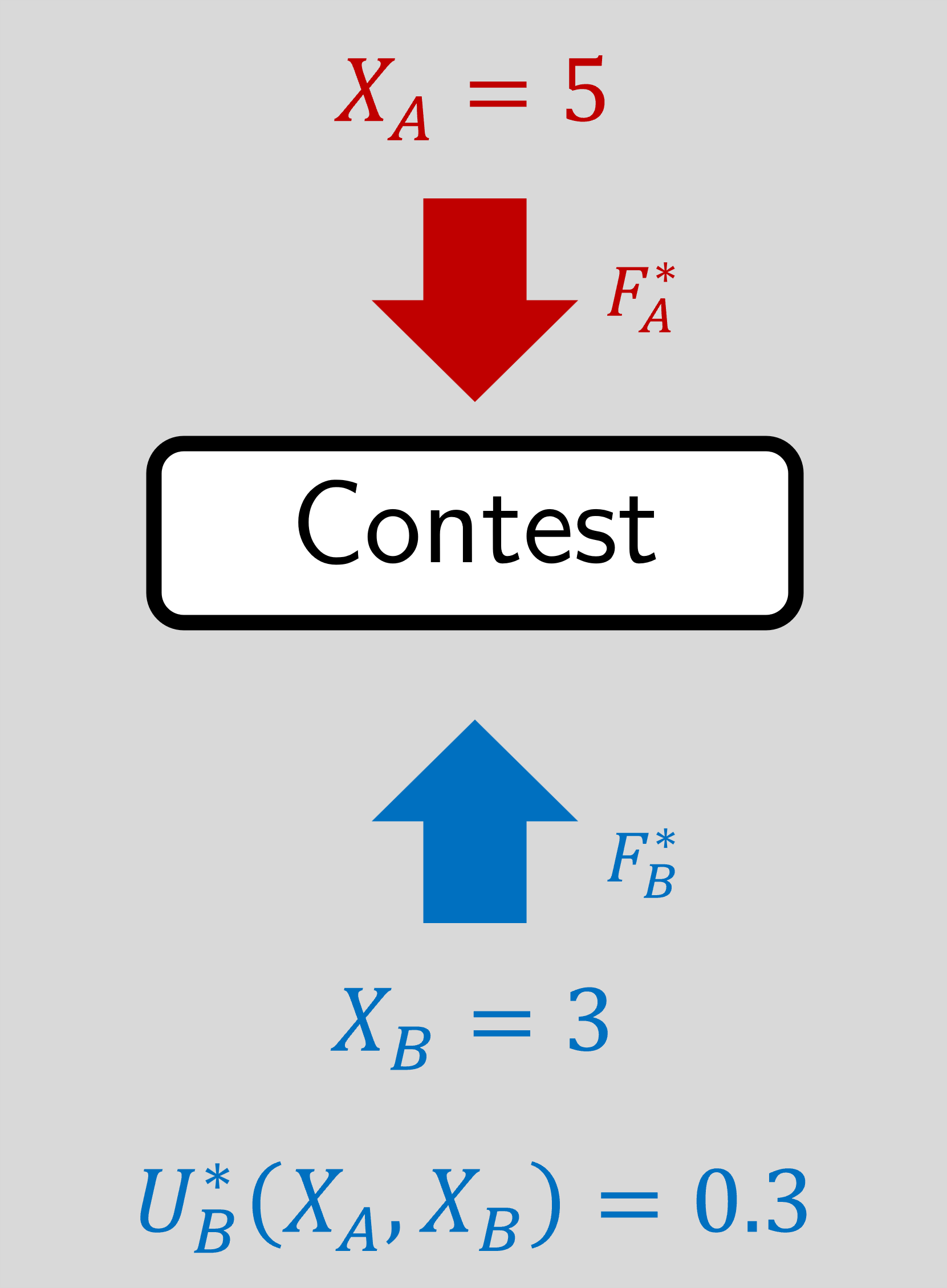}} \hfil
    \subfloat[ General Lotto game with Binary Information \label{fig:nex:gli}   ]{\includegraphics[height=20ex]{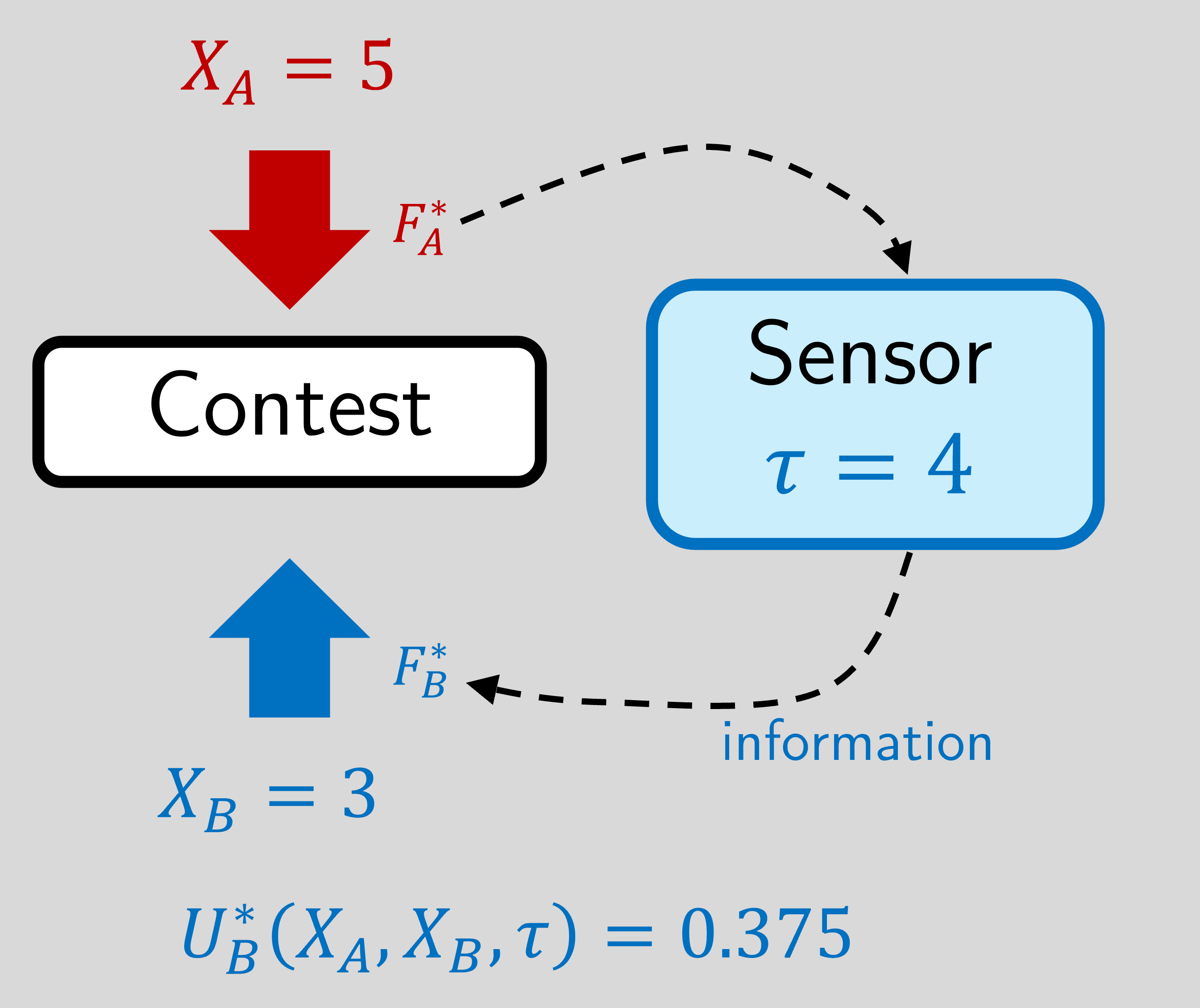}} 
    \caption{ Numerical instance of a General Lotto Game and a General Lotto Game with Binary Information. Note that placing a sensor that reveals if the realization is above or below $\tau=13$ generates a $\sim 13\%$ improvement to player $B$'s payoff.  }
    \label{fig:nex}
\end{figure}

In this section we present a numerical study to illustrate some relevant characteristics of the equilibrium payoff $U_B^*(X_A,X_B,\tau)$. Initially, let us illustrate how the partially unveiling the strategy of player $A$ suppose a strategic advantage for player $B$. With that in mind, we present the example depicted in Figure~\ref{fig:nex}.

In it, player $B$ with budget $X_B=10$, faces an adversary with budget $X_A=7$. When no informational advantage is given to player $B$, the expected utility is the usual General Lotto payoff as in Equation~\eqref{eq:lotto}, equal to $0.65$. On the other hand, with the binary information provided by the threshold $\tau$, player $B$ was able to increase its payoff to $0.7349$, improving by $\sim 13\%$. Thus, this numerical example illustrate the strategic opportunity provided by acquiring information, using a binary sensor, in a resource allocation scenario.

\begin{figure}[hbt]
    \centering
    \includegraphics[width=0.5\textwidth]{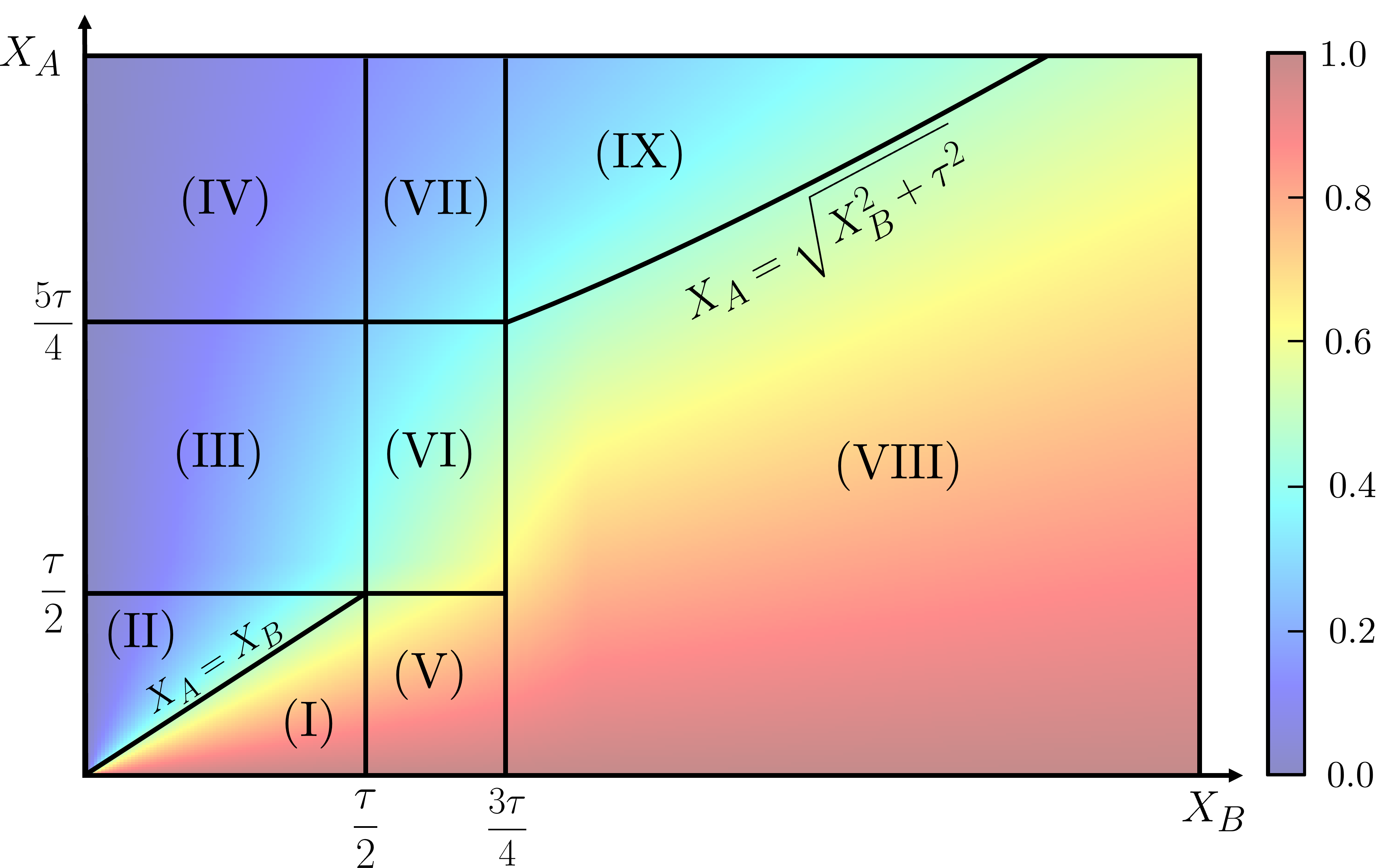}
    \caption{ Value of the function $U_B^*(X_A,X_B,\tau)$ over the regions described in Equations~\eqref{eq:ua:i}-\eqref{eq:ua:ix} in the $(X_A,X_B)$-plane.  }
    \label{fig:ua:reg}
\end{figure}

Now, we plot the function $U_B^*(X_A,X_B,\tau)$ for a fixed valued of $\tau$ as shown in Figure~\ref{fig:ua:reg}. We can observe the regions generated by the case statements in Equations~\eqref{eq:ua:i}-\eqref{eq:ua:ix}.

Note that, as player $B$ gets relatively stronger in term of its budget $X_B$, the payoff $U_B^*(X_A,X_B,\tau)$ increases. Then, including binary information in General Lotto games preserves the axiomatic property of contests of increasing the probability of winning a contest as the effort exerted increases. Also, if we set $\tau=0$, we can observe how most of the regions presented collapse at the $x$ and $y$ axis. Then, only two regions remain; $X_A\leq X_B$ (Equation~\eqref{eq:ua:viii}) and $X_B\leq X_A$ (Equation~\eqref{eq:ua:ix}). This confirm the fact that setting $\tau=0$ recover the original General Lotto formulation. Similarly, for $\tau$ sufficiently large ($\frac\tau2 \geq \max\crl{X_A,X_B}$) only two regions remain; $X_A\leq X_B$ (Equation~\eqref{eq:ua:i}) and $X_B\leq X_A$ (Equation~\eqref{eq:ua:ii}). Then, setting the value of the threshold extremely high also result in a regular General Lotto game. However, with our formulation we gain access to a characterization of all the possible cases in between.

\begin{figure}[hbt]
    \centering
    \includegraphics[width=0.5\textwidth]{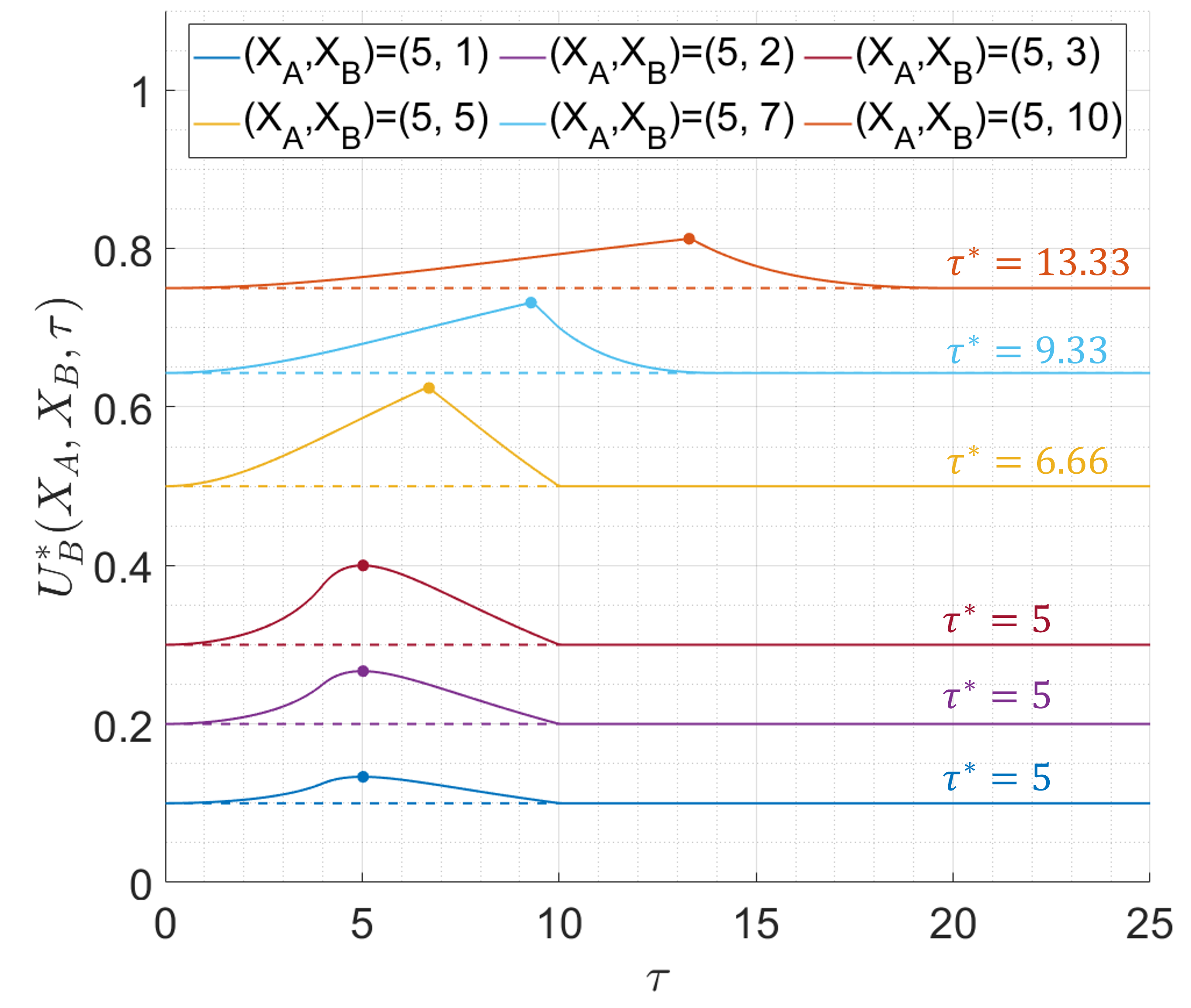}
    \caption{ Value of the function $U_B^*(X_A,X_B,\tau)$ as a function of $\tau$ for a different $(X_A,X_B)$ pairs. The dashed lines represent the payoff obtained by player $B$ in a General Lotto game $U_B^*(X_A,X_B)$. Also, we highlight the maximizer of each curve with a marker.  }
    \label{fig:tau}
\end{figure}

To analyze how the value of the threshold affects player $B$'s payoff we plot the value of $U_B^*(X_A,X_B,\tau)$ as a function of $\tau$ for different $(X_A,X_B)$ pairs. In addition, we plotted the obtained payoff in a General Lotto game to compare the payoff obtained with and without binary information. This plot is presented in Figure~\ref{fig:tau}.

As previously, we can observe that as player $B$ gets relatively stronger, the value of $U_B^*(X_A,X_B,\tau)$ increases. Also, note that the payoff for $\tau=0$ and for large enough $\tau$ match the value of the usual General Lotto game. Thus, setting $\tau$ to $0$ or to a large enough value just recover the payoff in a General Lotto game. Although, it can also be noticed that the value of $U_B^*(X_A,X_B,\tau)$ is always above or equal to the dashed line. This suggest that the information obtained by player $B$ is never hurtful; in the worst case, the obtained information is just useless. 

Moreover, the value of $U_B^*(X_A,X_B,\tau)$ achieves a maximum for certain value of $\tau$. This is highlighted by a dot for all the plots in Figure~\ref{fig:tau}. Then, the value of $\tau$ could be strategically chosen to maximize the benefit that information provides for player $B$. However, there is an intricate relationship between the player's endorsements, $X_A$ and $X_B$, and the optimal threshold.  

\section{Analysis}\label{sec:analysis}

In this section, we detail our technical approaches used to establish Theorem \ref{thm:info}. A high-level overview of our approach is given as follows.
\begin{itemize}
    \item First, we recognize that the binary signaling structure induces one of two possible sub-games to be played, which we call $\G_0$ or $\G_1$. They correspond to when the Breaker's received signal is $s=0$ or $s=1$.
    \item We derive equilibrium payoffs and strategies to each of the sub-games.
    \item The problem of deriving the equilibrium of $\GLI$ is equivalently cast as an  optimization problem for the Attacker $A$ (Lemma \ref{lm:equilibrium}).
    \item We solve the optimization problem to obtain the results reported in Theorem \ref{thm:info}.
\end{itemize}

\subsection{Definition of Sub-games}

Given the binary signaling structure of the threshold sensor, we may define the \emph{posterior distributions} of $F_A$ as
\begin{equation}
    F_{A|0}(x) := 
    \begin{cases}
        \frac{F_A(x)}{\Pr_{x\sim F_A}[x \leq \tau]}, &\text{if } x \leq \tau \\
        1, &\text{if } x > \tau
    \end{cases}
\end{equation}
\begin{equation}
    F_{A|1}(x) := 
    \begin{cases}
        0, &\text{if } x \leq \tau \\
        \frac{F_A(x) - F_A(\tau)}{\Pr_{x\sim F_A}[x > \tau]}, &\text{if } x > \tau \\
    \end{cases}
\end{equation}
In terms of the posterior distributions, the utility function \eqref{eq:GLI_utility} can equivalently be stated as
\begin{equation}\label{eq:GLI_utility_posterior}
    \begin{aligned}
        U_A(F_A,F_B,\tau) &= \Pr_{x\sim F_A}[x \leq \tau]\cdot U_A(F_{A|0},F_{B|0}) + \Pr_{x\sim F_A}[x > \tau]\cdot U_A(F_{A|1},F_{B|1})
    \end{aligned}
\end{equation}
where $U_A$ is defined in \eqref{eq:GL_utility}. Observe that $F_{A|0}$ is a distribution whose support is limited to the interval $[0,\tau)$. Likewise, $F_{A|1}$ is a distribution whose support is limited to the interval $[\tau,\infty)$. 
\begin{definition}
    Given an interval $[a,b)$, we say that a distribution $F$ belongs to $\I_{[a,b)}$ if its support is contained in $[a,b)$, i.e., $F \in \I_{[a,b)} \iff \supp\prth{ F } \subseteq [a,b)$.
\end{definition}
Consequently, we have that $F_{A|0} \in \I_{[0,\tau)}$ and $F_{A|1} \in \I_{[\tau,\infty)}$. We now observe that any $F_A \in \F(X_A)$ can be equivalently represented with three elements $(\alpha,F_{A|0},F_{A|1})$: a distribution $F_{A|0} \in \I_{[0,\tau)}$, a distribution $F_{A|1} \in \I_{[\tau,\infty)}$, and an $\alpha \in [0,1]$ such that together they satisfy,
\begin{equation}
    (1-\alpha) \Ex_{x\sim F_{A|0}}[x] + \alpha\Ex_{x\sim F_{A|1}}[x] = X_A.
\end{equation}
Here, $\alpha = \Pr_{x\sim F_A}[x > \tau]$ represents the probability mass from $F_A$ that is contained in the interval $(\tau,\infty)$.

Thus, player $A$ effectively selects a triple $(\alpha,F_{A|0},F_{A|1})$ as its strategy $F_A$. The expression in \eqref{eq:GLI_utility_posterior} suggests that player $A$ uses strategy $F_{A|0}$ in a sub-game induced when $s=0$ (occurring w.p. $1-\alpha$), and uses strategy $F_{A|1}$ in a sub-game induced when $s=1$ (occurring w.p. $\alpha$). We formally define these sub-games in the following definitions.

\begin{definition}[Upper Bounded General Lotto Game]
    The game $\G_0(X_0,X_B,\tau)$ is the simultaneous-move game where player $A$'s feasible strategies are all distributions $F_{A|0} \in \F(X_0) \cap \I_{[0,\tau)}$ where $X_0 < \tau$, and player $B$'s feasible strategies are all distributions $F_{B|0} \in \F(X_B)$. The payoff to player $A$ is $U_A(F_{A|0},F_{B|0})$.

    % Given a threshold $\tau$ and resource $X_0 < \tau$, the game $\G_0(X_0,X_B,\tau)$ is the one-shot game where player $A$'s feasible strategies are all distributions $F_{A|0} \in \F(X_0) \cap \I_{[0,\tau)}$, and player $B$'s feasible strategies are all distributions $F_{B|0} \in \F(X_B)$. The payoff to player $A$ is $U_A(F_{A|0},F_{B|0})$.
\end{definition}

\begin{definition}[Lower Bounded General Lotto Game]
    The game $\G_1(X_1,X_B,\tau)$ is the simultaneous-move game where player $A$'s feasible strategies are all distributions $F_{A|1} \in \F(X_1) \cap \I_{[\tau,\infty)}$ where $X_1 \geq \tau$, and player $B$'s feasible strategies are all distributions $F_{B|0} \in \F(X_B)$. The payoff to player $A$ is $U_A(F_{A|0},F_{B|0})$.

    % Given a threshold $\tau$ and resource $X_1 \geq \tau$ the game $\G_1(X_1,X_B,\tau)$ is the one-shot game where player $A$'s feasible strategies are all distributions $F_{A|1} \in \F(X_1) \cap \I_{[\tau,\infty)}$, and player $B$'s feasible strategies are all distributions $F_{B|1} \in \F(X_B)$. The payoff to player $A$ is $U_A(F_{A|1},F_{B|1})$.
\end{definition}

The sub-games $\G_0$ and $\G_1$ may be viewed as classic General Lotto games where player $A$'s strategy has a support constraint. In sub-game $\G_0$, player $A$ is constrained to choosing distributions with support on $[0,\tau)$, i.e. with an upper bound. In sub-game $\G_1$, player $A$ is constrained to choosing distributions with support on $[\tau,\infty)$, i.e. with a lower bound. Player $B$ has no such support constraints in either sub-game.

\subsection{Optimization problem to find equilibria of GLI}

In the lemma below, we establish that an equilibrium of $\GLI$ can be obtained by  solving a particular optimization problem.

\begin{lemma} \label{lm:equilibrium}
    The tuples $(\alpha^*, F_{A|0}^*,F_{A|1}^*)$ and $(F_{B|0}^*,F_{B|1}^*)$ constitute an equilibrium of $\GLI(X_A,X_B,\tau)$ if and only if
    \begin{enumerate}
        \item $(\alpha^*, F_{A|0}^*,F_{A|1}^*)$ is a solution to the optimization problem
        \begin{equation}\label{eq:MP}
            \begin{aligned}
                \max_{\alpha,F_{A|0},F_{A|1}} \quad & \biggl\{ (1-\alpha)U_A\prth{F_{A|0},F_{B|0}^*}  + \alpha U_A\prth{F_{A|1},F_{B|1}^*}  \biggr\}  \\
                \text{subject to }  & \alpha \in [0,1] \\
                                    & F_{A|0} \in \I_{[0,\tau)} \\
                                    & F_{A|1} \in \I_{[\tau,\infty)} \\
                                    & (1-\alpha) \Ex_{x\sim F_{A|0}}[x] + \alpha\Ex_{x\sim F_{A|1}}[x] = X_A. 
            \end{aligned} 
        \end{equation}
        
        \item $(F_{A|0}^*,F_{B|0}^*)$ is an equilibrium of $\G_0(\Ex_{x\sim F_{A|0}^*}[x],X_B,\tau)$.
        \item $(F_{A|1}^*,F_{B|1}^*)$ is an equilibrium of $\G_1(\Ex_{x\sim F_{A|1}^*}[x],X_B,\tau)$.
        \end{enumerate}
\end{lemma}

The proof is given in Appendix \ref{app:equivalent_OP}. In words, finding the equilibrium payoff to player $A$ in $\GLI$ (and consequently also to $B$) is equivalent to finding equilibrium profiles $\prth{F_{A|0}^*,F_{B|0}^*}$ and $\prth{F_{A|1}^*,F_{B|1}^*}$ to sub-games $\G_0$ and $\G_1$ respectively, and a weight $\alpha^* \in [0,1]$ such that the selected triple $(\alpha^*, F_{A|0}^*,F_{A|1}^*)$
\begin{itemize}
    \item  corresponds to an admissible $F_A \in \F(X_A)$ for the Attacker $A$.
    \item  maximizes player $A$'s expected payoff \eqref{eq:GLI_utility_posterior} given player $B$ responds with equilibrium strategies $F_{B|0}^*$, $F_{B|1}^*$ in each sub-game.
\end{itemize}
Lemma \ref{lm:equilibrium} provides a concrete path towards deriving solutions to $\GLI$. In particular,  we next need to characterize the equilibria to both sub-games $\G_0$ and $\G_1$ in isolation, in order for the optimization problem \eqref{eq:MP} to take an explicit formulation.

\subsection{Lower Bounded General Lotto Game Equilibrium}

Sub-game $\G_1(X_1,X_B,\tau)$ (for  $X_1 > \tau$) can be precisely mapped to a \emph{General Lotto game with Favoritism}, which has been studied in recent works~\cite{fav:vu,fav:chandan,fav:paarporn}. In a game with favoritism, one player (say $A$) has a ``head start" on the contest in the form of pre-allocated resources to the contest. In order to win, player $B$ needs to overcome the combined pre-allocated and usual resource allocation of player $A$.

% , say $\tau > 0$. Players are endowed with budgets $X_A,X_B$ to spend in the usual way, i.e. \eqref{eq:lotto_constraint}. A General Lotto game with favoritism is specified by a tuple $\text{GL-F}(X_A,X_B,\tau)$. to the game $\text{GL-F}(X_1 - \tau,X_B,\tau)$

Sub-game $\G_1(X_1,X_B,\tau)$ is identical to the General Lotto game with Favoritism where $A$ has head-start $\tau$ and remaining budget $X_1-\tau$, since $F_{A|1}$ has support in $[\tau,\infty)$ with expectation $X_1$. A complete characterization of its equilibria is given in the result below.

\begin{theorem}\label{thm:favoritism}
    \textbf{\emph{ (Adapted from Theorem~3.1 in~\cite{fav:paarporn} and Theorem~4.2 in~\cite{fav:vu}). }}
    Consider the game $\G_1(X_1,X_B,\tau)$ for some $X_1 > \tau$. Then player $A$'s (unique) equilibrium payoff $\pi^*_1(X_1,X_B,\tau)$ is given as follows:    
    \begin{enumerate}
        \item Suppose $X_B \leq \tau$, or $X_B > \tau$ and $X_1 - \tau \geq \frac{2(X_B-\tau)^2}{\tau+2(X_B-\tau)}$. Then, % the equilibrium payoff is given by,
        \begin{equation} \label{eq:case1_payoff}
            \pi^*_1(X_1,X_B,\tau) = 1 - \frac{X_B}{X_1 + \sqrt{X_1^2 - \tau^2}} .
        \end{equation}
        
        \item Suppose $X_B > \tau$ and $0 < X_1-\tau < \frac{2(X_B-\tau)^2}{\tau+2(X_B-\tau)}$. Then, % the equilibrium payoff is given by,
        \begin{equation} \label{eq:case2_payoff}
            \pi^*_1(X_1,X_B,\tau) = \frac{X_1 - \tau}{2(X_B-\tau)}.
        \end{equation}
        
        \item Suppose $X_1 = \tau$. Then, % the equilibrium profile is given by,
        \begin{equation} \label{eq:case3_payoff}
            \pi^*_1(X_1,X_B,\tau) = 1 - \min\crl{\frac{X_B}{\tau},1}.
        \end{equation}
    \end{enumerate}
\end{theorem}
With this characterization, the second term in the objective function in \eqref{eq:MP} can equivalently be replaced with $\pi^*_1(X_1,X_B,\tau)$, and the search variable $F_{A|1}\in\I_{[\tau,\infty)}$ becomes replaced with a scalar variable $X_1 \geq \tau$. It now remains to characterize equilibria to the other sub-game $\G_0$.

\subsection{Upper Bounded General Lotto Game Equilibrium}

We now turn our attention to deriving the equilibria of $\G_0(X_0,X_B,\tau)$. The full characterization is given in the Lemma below.

\begin{theorem} \label{thm:ub}
    Consider the game $\G_0(X_1,X_B,\tau)$ for some $X_0 \leq \tau$. Then player $A$'s equilibrium payoff $\pi^*_0(X_0,X_B,\tau)$ is given as follows:    
    \begin{enumerate}
        \item Suppose $X_B > \tau$. Then, % the equilibrium payoff is given by,
        \begin{equation} \label{eq:pi0:1}
            \pi^*_0(X_0,X_B,\tau) = 0
        \end{equation}
        
        \item Suppose $\tau \geq X_B \geq \frac\tau2 $. Then, % the equilibrium payoff is given by,
        \begin{equation} \label{eq:pi0:2} 
            \pi^*_0(X_0,X_B,\tau) = \begin{cases}
                \frac{2X_0}{\tau}\prth{ 1 - \frac{X_B}{\tau} },     & \text{ if $ 0 \leq X_0 \leq \frac\tau2$ } \\
                1 - \frac{X_B}{\tau},                               & \text{ if $ \frac\tau2 \leq X_0 \leq \tau$ }
            \end{cases}.
        \end{equation}
        
        \item Suppose $\frac\tau2 \geq X_B \geq 0$. Then, % the equilibrium payoff is given by,
        \begin{equation} \label{eq:pi0:3} 
            \pi^*_0(X_0,X_B,\tau) = \begin{cases}
                \frac{X_0}{2 X_B},      & \text{ if $ 0 \leq X_0 \leq X_B$ } \\
                1 - \frac{X_B}{2 X_0},  & \text{ if $ X_B \leq X_0 \leq \frac\tau2$ } \\
                1 - \frac{X_B}{\tau},   & \text{ if $ \frac\tau2  \leq X_0 \leq \tau$ } 
            \end{cases}.
        \end{equation}
    \end{enumerate}
\end{theorem}

Item 1 asserts that if the threshold $\tau$ is small enough ($\tau \leq X_B$), then player $B$ wins with total certainty. On the other hand, item 3 describes cases where if $\tau$ is larger than twice the budget of the stronger player, the equilibrium payoffs coincide with $\GL(X_0,X_B)$ (Theorem \ref{thm:GL}) because this does not constrain player $A$ from implementing its equilibrium strategy from the classic General Lotto game. 

In our proof, determining the equilibria for all other intermediate values of $\tau$ relies on leveraging upper bounded all-pay auctions (studied previously in~\cite[Table~1]{sensor}). However, it is worth mentioning here that directly leveraging  these methods provide equilibrium characterizations to only a subset of games $\G_0(X_0,X_B,\tau)$. In item 2 of Theorem \ref{thm:ub}, we fill these gaps to include all possible games $\G_0$. These proofs are found in the Appendix.

% These gaps, which we fill as novel contributions, suggest that player $A$ has equilibrium strategies that do not require utilizing its entire budget. \kp{These details and proofs can be found ???.}

\begin{figure}[hbt]
    \centering
    \subfloat[Case $X_0 \leq X_B$ \label{fig:case:0b}]{\includegraphics[width=0.42\textwidth]{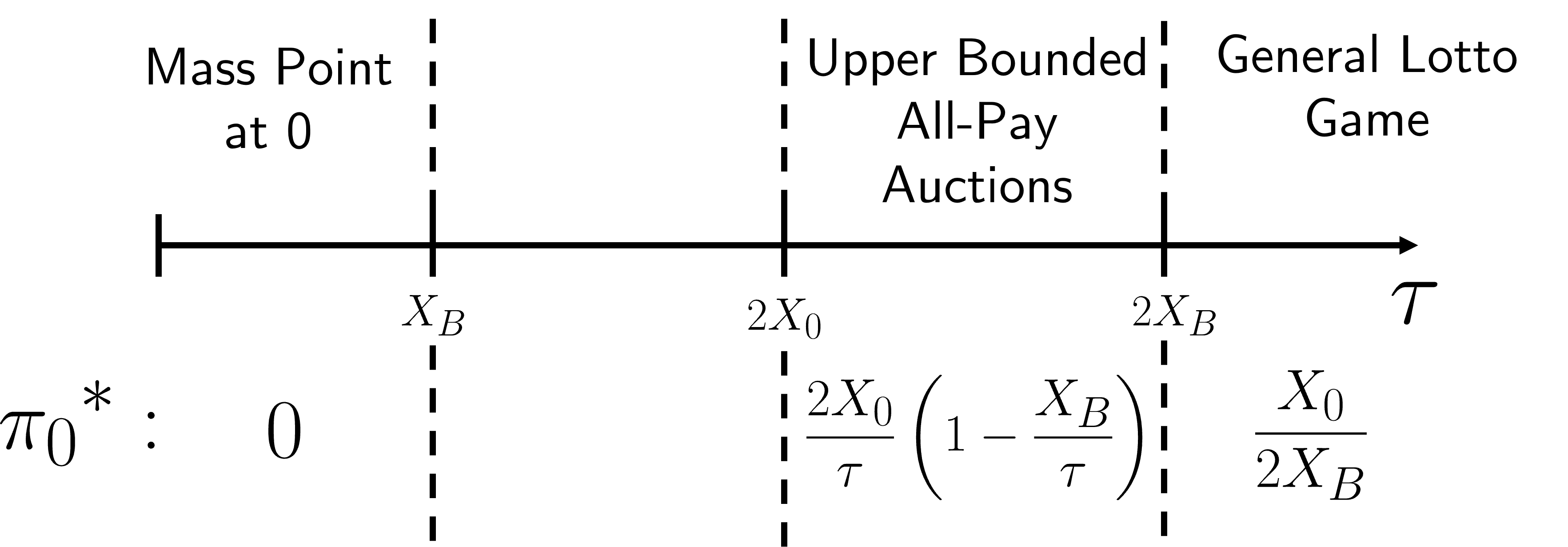}} \hfil
    \subfloat[Case $X_B \leq X_0$ \label{fig:case:b0}]{\includegraphics[width=0.42\textwidth]{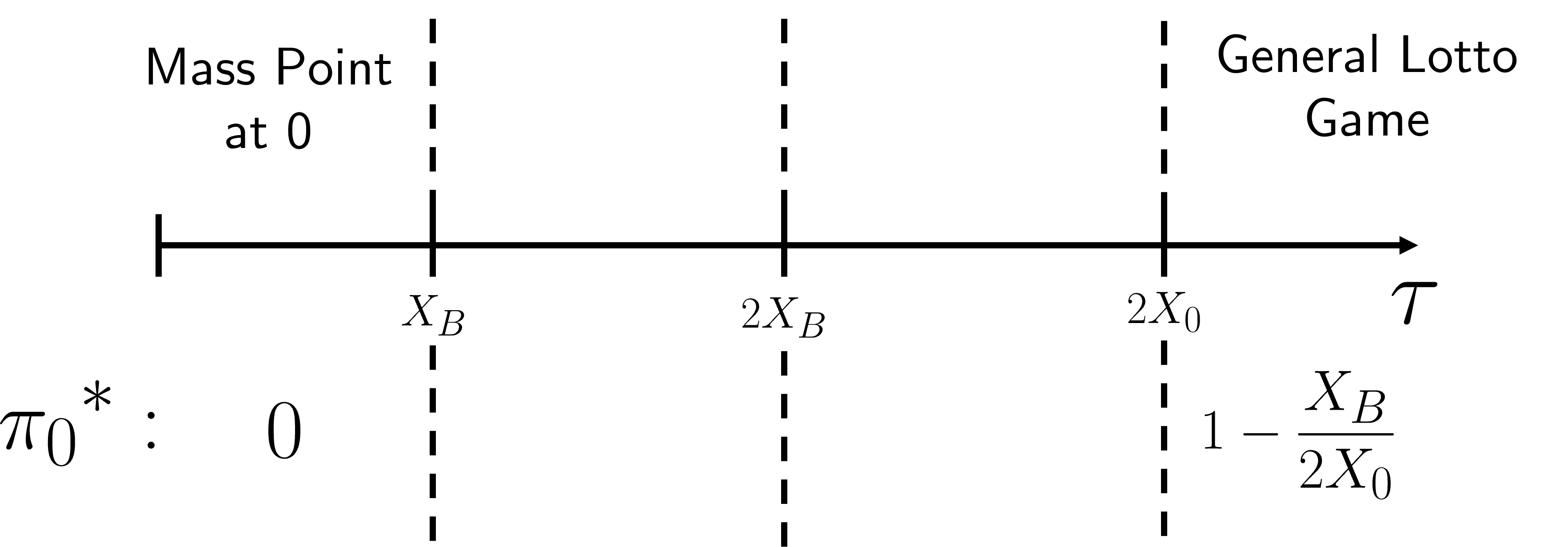}} 
    \caption{ Equilibrium payoff for the Upper Bounded General Lotto game. It can be noticed that, using previous results, do not allow us to have the equilibrium payoff for all possible values of $\tau$. }
    \label{fig:cases:ub}
\end{figure}

% Initially, to study such equilibrium we analyze how its relation to upper bounded all-pay auctions, previously studied in~\cite[Table~1]{sensor}. Although, establishing a relation between budgets in upper bounded Lotto games and values in upper bounded all-pay auctions do not allow to recover the entire $(X_0,X_B,\tau)$ space as shown in Figure~\ref{fig:cases:ub}.

% These gaps in the  $(X_0,X_B,\tau)$ space suggest that player $A$ must use equilibrium strategies that do not require its entire budget or, equivalently, $\Ex_{x\sim F_{A|0}}[x] < X_0$. For those regions, the equilibrium payoff is described in Lemma~\ref{lm:ub}.

\begin{lemma} \label{lm:ub}
    For the regions without equilibrium payoff in Figure~\ref{fig:cases:ub} we have that, $\pi_0^* = 1 - \frac{X_B}{\tau}$.
\end{lemma}

\subsection{General Lotto Games with Binary Information Equilibrium}

With the derivation of equilibrium payoffs for $\G_0(X_0,X_B,\tau)$ and $\G_1(X_1,X_B,\tau)$  (Theorems~\ref{thm:ub} and~\ref{thm:favoritism}) complete, we can rewrite the optimization problem presented from Lemma~\ref{lm:equilibrium} as follows.

\begin{corollary}
    The value of the optimization problem in~\eqref{eq:MP} is equivalent to the value of the following optimization problem,
    \begin{equation}\label{eq:MP2} 
    \begin{aligned}
        \max_{\alpha,X_0,X_1} &~\crl{ (1-\alpha)\pi_0^*\prth{X_0,X_B,\tau} + \alpha \pi_1^*\prth{ X_1, X_B,\tau} }  \\
        \text{subject to: } & \alpha \in [0,1] \\
                            & X_1 \geq \tau, \\
                            & (1-\alpha) X_0 + \alpha X_1 = X_A 
    \end{aligned} 
    \end{equation}
\end{corollary}

Therefore, the solution of the optimization problem in Equation~\eqref{eq:MP2} will give us the equilibrium payoff for player $A$ in $\GLI(X_A,X_B,\tau)$, which we denote $U_A^*(X_A,X_B,\tau)$. The equilibrium payoff for player $B$ can be obtained as $U_B^*(X_A,X_B,\tau) = 1 - U_A^*(X_A,X_B,\tau)$. Indeed, the results of our main Theorem \ref{thm:info} are derived precisely by solving \eqref{eq:MP2} for all possible parameter combinations $(X_A,X_B,\tau)$. The proofs are detailed in the Appendix \ref{sec:main_proof}.

\section{Conclusion}\label{sec:conclusion}

In this paper, we analyzed a General Lotto game where one of the players, termed the ``Breaker", is able to acquire information about the selected strategy of an opponent, termed the ``Attacker". The information is a signal that indicates whether the Attacker's effort was below or above a certain threshold.  Our main result comprehensively derives the equilibrium strategies and payoffs to both players. In numerical studies, the result demonstrates that the threshold sensor can significantly improve the Breaker's performance in comparison to the classic setting with no sensor. These contributions provide analytical tools and insights into evaluating the effectiveness of informationally-aware security policies. Future work is geared towards generalizing the level of granularity in the sensor signals, as well as deriving the optimal sensor thresholds.

% This scenario is representative of a defender that can unveil strategic intents of potential adversaries, and utilize the information to formulate a response.
%%%%%%%%%%%%%%%%%%%%%%%%%%%%%%%%%%%%%%%%%%%%%%%%%%%%%%%%%%%%%%%%%%%%%%%%%%%%%%%%%%%%%%%%%%%%%%%%%%%%%%%%%%%%%%%%%%%%%%%%%%%%%%%%%%%%%%%%%%%%%%%%%%%%%%%%%%%%%%%%%%%%%%%%%%%%%%%%%%%%%%%%%%%%%%%%%%%%%%%%%%%%%%%%%%%%%%%%%%%%%%%%%%%%%%%%%%%%%%%%%%%%%%%%%%%%%%%%%
\bibliographystyle{IEEEtran}
\bibliography{bib,kp_sources}

%%%%%%%%%%%%%%%%%%%%%%%%%%%%%%%%%%%%%%%%%%%%%%%%%%%%%%%%%%%%%%%%%%%%%%%%%%%%%%%%%%%%%%%%%%%%%%%%%%%%%%%%%%%%%%%%%%%%%%%%%%%%%%%%%%%%%%%%%%%%%%%%%%%%%%%%%%%%%%%%%%%%%%%%%%%%%%%%%%%%%%%%%%%%%%%%%%%%%%%%%%%%%%%%%%%%%%%%%%%%%%%%%%%%%%%%%%%%%%%%%%%%%%%%%%%%%%%%%
\appendix

%%%%%%%%%%%%%%%%%%%%%%%%%%%%%%%%%%%%%%%%%%%%%%%%%%%%%%%%%%%%%%%%%%%%%%%%%%%%%%%%%%%%%%%%%%%%%%%%%%%%%%%%%%%%%%%%%%%%%%%%%%%%%%%%%%%%%%%%%%%%%%%%%%%%%%%%%%%%%%%%%%%%%%%%%%%%%%%%%%%%%%%%%%%%%%%%%%%%%%%%%%%%%%%%%%%%%%%%%%%%%%%%%%%%%%%%%%%%%%%%%%%%%%%%%%%%%%%%%
\subsection{Proof of Lemma~\ref{lm:equilibrium}}\label{app:equivalent_OP}

The $\Rightarrow$ direction: item 1 is satisfied by definition of an equilibrium of GLI. For item 2, suppose the profile $(F_{A|0}^*,F_{B|0}^*)$ is not an equilibrium of $\G_0$. Then $A$ can deviate to another $F_{A|0}$ with the same expected value to improve its payoff. However, this would contradict item 1. Item 3 is verified in the same way.

The $\Leftarrow$ direction: Suppose items 1, 2, and 3 are satisfied. Then any deviation from player $A$ will not result in a strictly better payoff, due to item 1. Similarly, any deviation from player $B$ would not result in a strictly better payoff because of items 2 and 3.

\subsection{Proof of Lemma~\ref{lm:ub}}
There are three different regions without equilibrium characterization in Figure~\ref{fig:cases:ub}, \begin{itemize}
    \item $X_B \leq X_0$ and $2X_B \leq \tau \leq 2X_0$.
    \item $X_B \leq X_0$ and $X_B \leq \tau \leq 2X_B$.
    \item $X_0 \leq X_B$ and $X_B \leq 2X_0$.
\end{itemize}

Also, note that the regions with equilibrium payoff regions where $\tau \leq X_B$ player $A$ uses no budget. On the other hand, when $\tau \geq 2X_0$, the player $A$ uses all its budget $X_0$. This suggest that, in the regions in between, player $A$ does not use its entire budget. Let us define $\bar{X}_0$ as the used budget for player $A$. Now, we analyze each case individually.

\subsubsection{Case 1: $X_B \leq X_0$ and $2X_B \leq \tau \leq 2X_0$ }
Note that, even if player $A$ only uses $\frac\tau2$ units of budget, it still is the stronger player. That is, $X_B \leq \bar{X}_0 = \frac\tau2 \leq X_0$. Then, using the usual General Lotto equilibrium characterization, we obtain, 
\begin{equation*} \begin{aligned}
    F_A^*(x)    &= \frac{x}{2\bar{X}_0} = \frac{x}{\tau} \\
    F_B^*(x)    &= \prth{ 1 - \frac{X_B}{\bar{X}_0} } + \frac{X_B}{ 2\bar{X}_0^2 } x = \prth{ 1 - \frac{2X_B}{\tau} } + \frac{2X_B}{ \tau^2 } x \\
    \pi_A^*     &= 1 - \frac{X_B}{2\bar{X}_0} = 1 - \frac{X_B}{\tau}
\end{aligned} \end{equation*}

It is reasonable to think that player $A$ could improve its payoff by using all its budget. However, it can not allocate any resources above the threshold $\tau$. Then, we can characterize the possible strategies for player $A$ as a convex combination of $F_A^*(x)$ and some mass point in $\hat{x} \in [0,\tau]$. Then, the budget constraint translate into, 
\begin{equation*}
    \Ex[x] = \beta\prth{\frac{\tau}{2}} + (1-\beta)\hat{x},
\end{equation*}
and, the obtained payoff for player $A$ is, 
\begin{equation*} \begin{aligned}
    \pi_A   &= \beta\prth{1 - \frac{X_B}{\tau} } + (1-\beta) F_B\prth{ \hat{x} } \\
            &= \beta\prth{1 - \frac{X_B}{\tau} } + (1-\beta) \sqr{ \prth{1 - \frac{2X_B}{\tau} } + \frac{2X_B}{\tau^2}\hat{x} } \\
            &= \prth{1 - \frac{X_B}{\tau} } + (1-\beta)\frac{X_B}{\tau} \prth{ \frac{2\hat{x}}{\tau} - 1 }.
\end{aligned} \end{equation*}

Now, player $A$ could maximize its payoff by picking the appropriate mass point. This is equivalent to the maximization problem, 
\begin{equation*} \begin{aligned}
    \max_{\beta,\hat{x}} \qquad &\crl{ \prth{1 - \frac{X_B}{\tau} } + (1-\beta)\frac{X_B}{\tau} \prth{ \frac{2\hat{x}}{\tau} - 1 } }  \\
        \text{subject to: } & \beta \in [0,1] \\
                            & \hat{x} \in [0,\tau] \\
                            & \beta\prth{\frac{\tau}{2}} + (1-\beta)\hat{x} = X_0.
\end{aligned} \end{equation*}

If $\beta = 1$ then we recover the equilibrium where $\bar{X}_0 = \frac\tau2$. If $\beta \in [0,1)$ then, the equality constraint imply that, 
\begin{equation*}
    \hat{x} = \frac{2X_0 - \beta\tau}{2(1-\beta)}
\end{equation*}
and then, 
\begin{align*}
    (1-\beta)\frac{X_B}{\tau} \prth{ \frac{2\hat{x}}{\tau} - 1 }    &=  (1-\beta)\frac{X_B}{\tau} \sqr{ \frac{2}{\tau} \prth{ \frac{2X_0 - \beta\tau}{2(1-\beta)} } - 1 } \\
                                                                    &= \frac{X_B}{\tau}\prth{ \frac{2X_0 - \tau}{\tau} }.
\end{align*}

Since tie breakers are given to player $B$, player $A$ chooses to randomize uniformly over $[0,\tau]$ since including a mass point does not generate any profit. Then, 
\begin{equation*}
   \pi_A^* = 1- \frac{X_B}{\tau}. 
\end{equation*}

\subsubsection{Case 2: $X_B \leq X_0$ and $X_B \leq \tau \leq 2X_B$ }
Similarly, to the previous case, player $A$ chooses to uses budget equal to $\bar{X}_0 = \frac\tau2$. However, that means that player $A$ is not the stronger player anymore. Given that player $B$ does not have to constraint is support, it can use a strategy that randomizes uniformly over $[0,\tau]$ with probability $\gamma$ and a mass point in $\tau+\eps$ with probability $1-\gamma$. This mass point ensures that player $B$ wins with probability $1$. Therefore, player $A$ only can win as a consequence of the randomization that occurs with probability $\gamma$. 

Similar as before, player $A$ randomizes uniformly over $[0,\tau]$ with probability $\beta$ and put some mass point at $\hat{x}$ with probability $1-\beta$. The budget constraint implies that, 
\begin{equation*}
    \hat{x} = \frac{2X_0 - \beta\tau}{2(1-\beta)}. 
\end{equation*}

And, the payoff for player $A$ is, 
\begin{align*}
    \pi_A   &= \gamma\sqr{ \beta \prth{\frac12} + (1-\beta)F_B(\hat{x}) }
            = \gamma\sqr{ \beta \prth{\frac12} + (1-\beta) \frac{\hat{x}}{\tau} } \\
            &= \gamma\sqr{ \beta \prth{\frac12} +  \frac{(1-\beta)}{\tau}\frac{2X_0 - \beta\tau}{2(1-\beta)} } 
            = \gamma\frac{X_0}{\tau}.
\end{align*}

Again, player $A$ chooses to randomize uniformly over $[0,\tau]$ since including a mass point does not generate any profit. To find the value of $\gamma$ player $B$ has to satisfy its budget constraint, 
\begin{align*}
            X_B     = \gamma\prth{\frac{\tau}{2} } + (1-\gamma)(\tau+\eps) 
    % \iff    X_B - (\tau+\eps) &= \gamma\sqr{ \frac\tau2 - (\tau+\eps) } \\
    \iff    \gamma = \frac{2( \tau  + \eps - X_B )}{\tau + 2\eps}.
\end{align*}

Given that tie-breakers are given to player $B$, it can set $\eps \downarrow 0$. Then, 
\begin{align*}
    \pi_A = \frac{\gamma}{2} = \frac{\tau  + \eps - X_B}{\tau + 2\eps} \xRightarrow{\eps \downarrow 0} \pi_A^* = 1 - \frac{X_B}{\tau}.
\end{align*}

\subsubsection{Case 3: $X_0 \leq X_B$ and $X_B \leq 2X_0$ }
Using the same reasoning, we assume that $\bar{X}_0 = \frac\tau2 \leq X_B$. Since player $A$ is the weaker player in this scenario, the equilibrium payoff will be the one obtained in Case 2, where player $A$ also was the weaker player. Therefore, 
\begin{align*}
    \pi_A^* = 1 - \frac{X_B}{\tau}.
\end{align*}

%%%%%%%%%%%%%%%%%%%%%%%%%%%%%%%%%%%%%%%%%%%%%%%%%%%%%%%%%%%%%%%%%%%%%%%%%%%%%%%%%%%%%%%%%%%%%%%%%%%%%%%%%%%%%%%%%%%%%%%%%%%%%%%%%%%%%%%%%%%%%%%%%%%%%%%%%%%%%%%%%%%%%%%%%%%%%%%%%%%%%%%%%%%%%%%%%%%%%%%%%%%%%%%%%%%%%%%%%%%%%%%%%%%%%%%%%%%%%%%%%%%%%%%%%%%%%%%%%
\subsection{Proof of Theorem~\ref{thm:info}}\label{sec:main_proof}
To find player's $A$ equilibrium payoff we need to solve the optimization problem presented in Equation~\eqref{eq:MP2}. With that in mind, we use the definitions of the equilibrium payoffs in in Theorems~\ref{thm:favoritism} and~\ref{thm:ub}. This can be rewritten as the following optimization problem, 
\begin{equation} \label{eq:optProb} \begin{aligned}
     U_A^*(X_A,X_B,\tau) = \max_{\substack{ X_1 \geq \tau \\ X_1 \geq X_A} } \crl{ \max_{ \alpha \in \sqr{0,\frac{X_A}{X_1}} } \sqr{ (1-\alpha) \pi_0^*\prth{ \frac{X_A-\alpha X_1}{1-\alpha}, X_B, \tau } + \alpha \pi_1^*\prth{ X_1, X_B, \tau } } }.
\end{aligned} \end{equation}

Since both $\pi_0^*(X_0,X_B,\tau)$ and $\pi_1^*(X_1,X_B,\tau)$ have multiple case statements a solution for each optimization problem must be provided. In Table~\ref{tab:cases} we list all the possible optimization problems that emerge from the cases statements in Equations~\eqref{eq:case1_payoff}-\eqref{eq:case3_payoff} and Equations~\eqref{eq:pi0:1}-\eqref{eq:pi0:3}. 
\begin{table}[htb]
\caption{Multiple cases for the optimization problem in Equation~\eqref{eq:optProb}}
\centering
\begin{tabular}{|llll|}
\hline
\multicolumn{4}{|l|}{\#1: $X_B>\tau$} \\ \hline
\multicolumn{1}{|l|}{\multirow{8}{*}{ \makecell{ \#2: \\ $X_B \leq \tau$} }} & \multicolumn{3}{l|}{\#2.1: $\frac\tau2 \leq X_0$} \\[1.1ex] \cline{2-4} 
\multicolumn{1}{|l|}{}                     & \multicolumn{1}{l|}{\multirow{2}{*}{ \makecell{ \#2.2: \\ $X_B \leq X_0 \leq \frac\tau2$} }} & \multicolumn{2}{l|}{\#2.2.1: $X_A\leq \frac\tau2$ } \\ \cline{3-4} 
\multicolumn{1}{|l|}{}                     & \multicolumn{1}{l|}{}                       & \multicolumn{2}{l|}{\#2.2.2: $X_A\geq \frac\tau2$} \\ \cline{2-4} 
\multicolumn{1}{|l|}{}                     & \multicolumn{1}{l|}{\multirow{3}{*}{ \makecell{ \#2.3: \\ $X_0 \leq \frac\tau2 \leq X_B$} }} & \multicolumn{1}{l|}{\multirow{2}{*}{ \makecell{ \#2.3.1: \\ $\frac\tau2 \leq X_B \leq \frac34\tau$} }} & \makecell{ \#2.3.1.1: \\ $X_A \leq \frac\tau2$} \\ \cline{4-4} 
\multicolumn{1}{|l|}{}                     & \multicolumn{1}{l|}{}                       & \multicolumn{1}{l|}{}                         & \makecell{ \#2.3.1.2: \\ $X_A \geq \frac\tau2$ } \\ \cline{3-4} 
\multicolumn{1}{|l|}{}                     & \multicolumn{1}{l|}{}                       & \multicolumn{2}{l|}{\#2.3.2: $\frac34 \tau \leq X_B \leq \tau$}                  \\ \cline{2-4} 
\multicolumn{1}{|l|}{}                     & \multicolumn{1}{l|}{\multirow{2}{*}{ \makecell{ \#2.4: \\ $X_0 \leq X_B \leq \frac\tau2$} }} & \multicolumn{2}{l|}{\#2.4.1: $X_A \leq X_B$}                                     \\ \cline{3-4} 
\multicolumn{1}{|l|}{}                     & \multicolumn{1}{l|}{}                       & \multicolumn{2}{l|}{\#2.4.2: $X_B \leq X_A$}                                     \\ \hline
\end{tabular}
\label{tab:cases}
\end{table}

Since both $\pi_0^*(X_0,X_B,\tau)$ and $\pi_1^*(X_1,X_B,\tau)$ have multiple case statements, we present the solution for each one of the possible cases. 

\subsubsection{Case \#1: $X_B > \tau$ } 
For all the cases when $X_B > \tau$ we have that $\pi_0^*(X_0,X_B,\tau) = 0$. Then, the optimization problem in Equation~\eqref{eq:optProb} can be simplified as, 
\begin{align*}
    & \max_{\substack{ X_1 \geq \tau \\ X_1 \geq X_A} } \crl{ \max_{ \alpha \in \sqr{0,\frac{X_A}{X_1}} } \sqr{ (1-\alpha) \pi_0^*\prth{ \frac{X_A-\alpha X_1}{1-\alpha}, X_B, \tau } + \alpha \pi_1^*\prth{ X_1, X_B, \tau } } } \\
    = & \max_{\substack{ X_1 \geq \tau \\ X_1 \geq X_A} } \crl{ \max_{ \alpha \in \sqr{0,\frac{X_A}{X_1}} } \sqr{ \alpha \pi_1^*\prth{ X_1, X_B, \tau } } }
\end{align*}
Since $\pi_1^*(X_1,X_B,\tau) \geq 0$ and its construction ensures that $X_1\geq\tau$, the optimization problem can be written as, 
\begin{align*}
    \max_{ X_1 \geq X_A } \crl{ \frac{X_A}{X_1} \pi_1^*(X_1,X_B,\tau) }
\end{align*}

There are 3 possible definitions for the function $\frac{X_A}{X_1} \pi_1^*(X_1,X_B,\tau)$, 
\begin{align*}
    \frac{X_A}{X_1} \pi_1^*(X_1,X_B,\tau) = \begin{cases}
        \frac{X_A}{X_1} \sqr{ 1 - \min\crl{ \frac{X_B}{\tau}, 1 } }             & \text{ if } X_1 = \tau \\
        \frac{X_A}{X_1} \sqr{ \frac{X_1 - \tau}{2(X_B-\tau)} }                  & \text{ if } \tau \leq X_1 \leq \frac{2(X_B-\tau)^2}{\tau + 2(X_B-\tau)} + \tau \\
        \frac{X_A}{X_1} \sqr{ 1 - \frac{X_B}{X_1 + \sqrt{ X_1^2 - \tau^2 }} }   & \text{ if } \leq \frac{2(X_B-\tau)^2}{\tau + 2(X_B-\tau)} + \tau \leq X_1
    \end{cases}
\end{align*}

Since we are assuming that $X_B > \tau$ then, 
\begin{align*}
    X_B > \tau  & \iff \min\crl{ \frac{X_B}{\tau} , 1 } = 1 \\
                & \iff \frac{X_A}{X_1} \sqr{ 1 - \min\crl{ \frac{X_B}{\tau}, 1 } }  = 0.
\end{align*}

Also, note that the function, 
\begin{align*}
    \frac{X_A}{X_1} \sqr{ \frac{X_1 - \tau}{2(X_B-\tau)} } = \frac{X_A}{2(X_B-\tau)} - \frac{X_A\tau}{2(X_B-\tau)} \frac{1}{X_1},
\end{align*}
is an increasing function of $X_1$. Finally, note that, 
\begin{align*}
    \frac{\partial}{\partial X_1} \crl{ \frac{X_A}{X_1} \sqr{ 1 - \frac{X_B}{X_1 + \sqrt{ X_1^2 - \tau^2 }} } } = -\frac{X_A}{X_1^2} \sqr{ 1 - \frac{X_B}{ \sqrt{ X_1^2 - \tau^2 } } }.
\end{align*}

Therefore, the function $\frac{X_A}{X_1} \pi_1^*(X_1,X_B,\tau)$ is an increasing continuous function of $X_1$ for $X_1 \leq \sqrt{ X_B^2 + \tau^2 }$ and decreasing for $X_1 \geq \sqrt{ X_B^2 + \tau^2 }$. Then, the value $\sqrt{ X_B^2 + \tau^2 }$ is a maximizer of $\frac{X_A}{X_1} \pi_1^*(X_1,X_B,\tau)$. With the constraint $X_1 \geq X_A$ we obtain that, 
\begin{align*}
    U_A^*(X_A,X_B,\tau) =  \begin{cases}
        \frac{X_A}{X_B + \sqrt{ X_B^2 + \tau^2 } }             & \text{ if } \sqrt{X_B^2+\tau^2} \geq X_A \\
        1 - \frac{X_B}{X_A + \sqrt{X_A^2-\tau^2}}              & \text{ if } \sqrt{X_B^2+\tau^2} \leq X_A
    \end{cases}
\end{align*}

\subsubsection{Case \#2: $X_B \leq \tau$ } 
When $X_B \leq \tau$ the value of $\pi_1^*(X_1,X_B,\tau)$ is always equal to $1 - \frac{X_B}{X_1 + \sqrt{ X_1^2 - \tau^2 }}$. Then, Equation~\eqref{eq:optProb} can be simplified as, 
\begin{align*}
    U_A^*(X_A,X_B,\tau) = & \max_{\substack{ X_1 \geq \tau \\ X_1 \geq X_A} } \crl{ \max_{ \alpha \in \sqr{0,\frac{X_A}{X_1}} } \sqr{ (1-\alpha) \pi_0^*\prth{ \frac{X_A-\alpha X_1}{1-\alpha}, X_B, \tau } + \alpha \pi_1^*\prth{ X_1, X_B, \tau } } } \\
    = & \max_{\substack{ X_1 \geq \tau \\ X_1 \geq X_A} } \crl{ \max_{ \alpha \in \sqr{0,\frac{X_A}{X_1}} } \sqr{ (1-\alpha) \pi_0^*\prth{ \frac{X_A-\alpha X_1}{1-\alpha}, X_B, \tau } + \alpha \prth{ 1 - \frac{X_B}{X_1 + \sqrt{ X_1^2 - \tau^2 }} } } }
\end{align*}

\subsubsection{Case \#2.1: $X_0 \geq \frac\tau2$ } 
\begin{align*}
    U_A^*(X_A,X_B,\tau) = & \max_{\substack{ X_1 \geq \tau \\ X_1 \geq X_A} } \crl{ \max_{ \alpha \in \sqr{0,\frac{X_A}{X_1}} } \sqr{ (1-\alpha) \prth{ 1 - \frac{X_B}{\tau} } + \alpha \prth{ 1 - \frac{X_B}{X_1 + \sqrt{ X_1^2 - \tau^2 }} } } } \\
    = & \max_{\substack{ X_1 \geq \tau \\ X_1 \geq X_A} } \crl{ \max_{ \alpha \in \sqr{0,\frac{X_A}{X_1}} } \sqr{ \prth{ 1 - \frac{X_B}{\tau} } + \alpha X_B \prth{ \frac{1}{\tau} - \frac{1}{X_1 + \sqrt{ X_1^2 - \tau^2 }} } } }
\end{align*}

Note that the constraint $X_0 \geq \frac\tau2$ implies that, 
\begin{align*}
    X_0 = \frac{X_A - \alpha X_1}{1-\alpha} \geq \frac\tau2 \iff \alpha \leq \frac{X_A - \frac\tau2}{X_1 - \frac\tau2}.
\end{align*}

Therefore, it is required that $ X_A \geq \frac\tau2$. Then, 
\begin{align*}
    U_A^*(X_A,X_B,\tau) = & \max_{\substack{ X_1 \geq \tau \\ X_1 \geq X_A} } \crl{ \max_{ \alpha \in \sqr{0,\frac{X_A - \frac\tau2}{X_1 - \frac\tau2} } } \sqr{ \prth{ 1 - \frac{X_B}{\tau} } + \alpha X_B \prth{ \frac{1}{\tau} - \frac{1}{X_1 + \sqrt{ X_1^2 - \tau^2 }} } } }.
\end{align*}

Since, 
\begin{align*}
    X_1 \geq \tau   & \implies X_1 + \sqrt{ X_1^2 - \tau^2 } \geq \tau \\
                    & \implies \frac{1}{\tau} - \frac{1}{X_1 + \sqrt{ X_1^2 - \tau^2 }} \geq 0, 
\end{align*}
we can ensure that $\alpha^* = \frac{X_A - \frac\tau2}{X_1 - \frac\tau2}$. Thus, 
\begin{align*}
    U_A^*(X_A,X_B,\tau) = & \max_{\substack{ X_1 \geq \tau \\ X_1 \geq X_A} } \crl{ \prth{ 1 - \frac{X_B}{\tau} } + \frac{2 X_B}{\tau} \prth{ X_A - \frac\tau2 } \frac{1}{2X_1 - \tau} \prth{ 
1 - \frac{\tau}{X_1 + \sqrt{X_1^2 - \tau^2} } } }.
\end{align*}

Now, 
\begin{align*}
    \frac{\partial}{\partial X_1} \sqr{ \frac{1}{2X_1 - \tau} \prth{ 1 - \frac{\tau}{X_1 + \sqrt{X_1^2 - \tau^2} } } } = \frac{\tau(2X_1 - \tau) - 2\sqrt{X_1^2-\tau^2}( X_1 + \sqrt{X_1^2 -\tau^2} - \tau )}{ (2X_1-\tau)^2 ( X_1 + \sqrt{X_1^2 - \tau^2} ) \sqrt{X_1^2 - \tau^2} }.
\end{align*}

Then, by first order conditions, 
\begin{align*}
    \tau(2X_1 - \tau) - 2\sqrt{X_1^2-\tau^2}\prth{ X_1 + \sqrt{X_1^2 -\tau^2} - \tau } = 0 \implies X_1 = \frac54 \tau.
\end{align*}

Therefore, 
\begin{align*}
    X_1^* = \max\crl{ \frac54 \tau, X_A }.
\end{align*}

Equivalently, for $X_B\leq \tau$ and $X_A\geq \frac\tau2$, we have that, 
\begin{align*}
    U_A^*(X_A,X_B,\tau) =  \begin{cases}
        \prth{ 1 - \frac{X_B}{\tau} } + \frac{X_B(2X_A - \tau)}{3\tau^2}    & \text{ if } X_A \leq \frac54\tau \\
        1 - \frac{X_B}{X_A + \sqrt{X_A^2-\tau^2}}                           & \text{ if } X_A \geq \frac54\tau
    \end{cases}
\end{align*}

\subsubsection{Case \#2.2: $X_B \leq X_0 \leq \frac\tau2$ } 
\begin{align*}
    U_A^*(X_A,X_B,\tau) = & \max_{\substack{ X_1 \geq \tau \\ X_1 \geq X_A} } \crl{ \max_{ \alpha \in \sqr{0,\frac{X_A}{X_1}} } \sqr{ (1-\alpha) \sqr{ 1 - \frac{X_B}{2}\prth{ \frac{1-\alpha}{X_A - \alpha X_1} } } + \alpha \prth{ 1 - \frac{X_B}{X_1 + \sqrt{ X_1^2 - \tau^2 }} } } } \\
    = & \max_{\substack{ X_1 \geq \tau \\ X_1 \geq X_A} } \crl{ \max_{ \alpha \in \sqr{0,\frac{X_A}{X_1}} } \crl{  1 - \frac{X_B}{2(X_1 + \sqrt{X_1^2 - \tau^2})} \sqr{ \prth{ X_1 + \sqrt{X_1^2-\tau^2}} \frac{(1-\alpha)^2}{(X_A - \alpha X_1)} + 2\alpha } } }
\end{align*}

Then, the conditions on $X_0$ imply that, 
\begin{align*}
    X_B \leq X_0 = \frac{X_A - \alpha X_1}{1-\alpha} \leq \frac\tau2 \iff \frac{X_A - \frac\tau2}{X_1 - \frac\tau2} \leq \alpha \leq \frac{X_A - X_B}{X_1 - X_B}.
\end{align*}

For the inner optimization problem we have,
\begin{align*}
    \frac{\partial}{\partial \alpha} \sqr{ \prth{ X_1 + \sqrt{X_1^2-\tau^2}} \frac{(1-\alpha)^2}{(X_A - \alpha X_1)} + 2\alpha } = \frac{ (X_1 + \sqrt{X_1^2-\tau^2})(1-\alpha) \sqr{ X_1(1-\alpha) + 2(\alpha X_1 - X_A) } + 2(X_A - \alpha X_1)^2 }{ ( X_A - \alpha X_1)^2 }
\end{align*}
By first order conditions, 

\resizebox{\hsize}{!}{
\begin{minipage}{0.95\linewidth}
\begin{align*}
    \prth{X_1 + \sqrt{X_1^2-\tau^2}}(1-\alpha) \sqr{ X_1(1-\alpha) + 2(\alpha X_1 - X_A) } + 2(X_A - \alpha X_1)^2 &= 0 \\
    \alpha^2 \sqr{ 2X_1^2 - X_1\prth{ X_1 + \sqrt{X_1^2-\tau^2} } } + \alpha \sqr{ 2X_A\prth{ X_1 + \sqrt{X_1^2-\tau^2} - 4X_1X_A } } + \sqr{ (X_1 - 2X_A)\prth{ X_1 + \sqrt{X_1^2-\tau^2} } + 2X_A^2 } &= 0
\end{align*}
\end{minipage} }

Note, by using the discriminant of a quadratic function, that the equation above does not have real solutions. Then, $\prth{ X_1 + \sqrt{X_1^2-\tau^2}} \frac{(1-\alpha)^2}{(X_A - \alpha X_1)} + 2\alpha$ is a monotone function of $\alpha$. Furthermore, by checking the vertex of the quadratic function, we can ensure that it is an increasing function of $\alpha$. Then, 
\begin{align*}
    \alpha^* = \max\crl{ 0, \frac{X_A - \frac\tau2}{X_1 - \frac\tau2} }
\end{align*}

\subsubsection{Case \#2.2.1: $X_A \leq \frac\tau2$ } 
To have a non-empty domain for the inner optimization problem, it is required that, 
\begin{align*}
    0 \leq \frac{X_A - X_B}{X_1 - X_B} \iff X_A \geq X_B. 
\end{align*}

Therefore, $\alpha^* = 0$ and, 
\begin{align*}
    U_A^*(X_A,X_B,\tau) &= \max_{\substack{ X_1 \geq \tau \\ X_1 \geq X_A} } \crl{ \max_{ \alpha \in \sqr{0,\frac{X_A}{X_1}} } \crl{  1 - \frac{X_B}{2(X_1 + \sqrt{X_1^2 - \tau^2})} \sqr{ \prth{ X_1 + \sqrt{X_1^2-\tau^2}} \frac{(1-\alpha)^2}{(X_A - \alpha X_1)} + 2\alpha } } } \\
    &= \max_{\substack{ X_1 \geq \tau \\ X_1 \geq X_A} } \crl{  1 - \frac{X_B}{2X_A} }  =  1 - \frac{X_B}{2X_A} 
\end{align*}

\subsubsection{Case \#2.2.2: $X_A \geq \frac\tau2$ } 
To have a non-empty domain for the inner optimization problem, it is required that, 
\begin{align*}
    \frac{X_A - \frac\tau2}{X_1 - \frac\tau2} \leq \frac{X_A - X_B}{X_1 - X_B} \iff X_B \leq \frac\tau2,  
\end{align*}
which is true by assumption. Therefore, 
\begin{align*}
    X_1^* &= \argmin_{X_1 \geq X_A} \frac{(1-\alpha)^2}{(X_A - \alpha X_1)} + \frac{2\alpha}{X_1 + \sqrt{X_1^2-\tau^2}} \qquad \text{ with } \alpha = \frac{X_A - \frac\tau2}{X_1 - \frac\tau2} \\
    &= \argmin_{X_1 \geq X_A} \frac{2\prth{X_A - \frac\tau2}}{X_1 - \frac\tau2} \sqr{ \frac{X_1 - X_A}{\tau\prth{ X_A - \frac\tau2 }} + \frac{1}{X_1 + \sqrt{X_1^2-\tau^2} } }.
\end{align*}

Now, 
\begin{align*}
    & \frac{\partial}{\partial X_1} \crl{ \frac{2\prth{X_A - \frac\tau2}}{X_1 - \frac\tau2} \sqr{ \frac{X_1 - X_A}{\tau\prth{ X_A - \frac\tau2 }} + \frac{1}{X_1 + \sqrt{X_1^2-\tau^2} } } } \\
    & = \frac{2\prth{X_A - \frac\tau2}}{\prth{X_1 - \frac\tau2}^2 } \sqr{ \frac{ \sqrt{X_1^2 - \tau^2} \prth{ X_1 + \sqrt{X_1^2-\tau^2} } - \tau\prth{ X_1 - \frac\tau2 + \sqrt{X_1^2-\tau^2} } }{ \tau \sqrt{X_1^2 - \tau^2} \prth{X_1 + \sqrt{X_1^2 - \tau^2} } } }
\end{align*}

By first order conditions, 
\begin{align*}
    \sqrt{X_1^2 - \tau^2} \prth{ X_1 + \sqrt{X_1^2-\tau^2} } - \tau\prth{ X_1 - \frac\tau2 + \sqrt{X_1^2-\tau^2} } = 0 
    \implies X_1 = \frac54 \tau.
\end{align*}

Therefore, 
\begin{align*}
    X_1^* = \max\crl{ \frac54 \tau, X_A }.
\end{align*}

In summary, for $X_B \leq \frac\tau2$ and $X_A \geq \frac\tau2$ we have,
\begin{align*}
    U_A^*(X_A,X_B,\tau) =  \begin{cases}
        \prth{ 1 - \frac{X_B}{\tau} } + \frac{X_B(2X_A - \tau)}{3\tau^2}    & \text{ if } X_A \leq \frac54\tau \\
        1 - \frac{X_B}{X_A + \sqrt{X_A^2-\tau^2}}                           & \text{ if } X_A \geq \frac54\tau
    \end{cases},
\end{align*}
and for $X_B \leq X_A \leq \frac\tau2$ we have, 
\begin{align*}
    U_A^*(X_A,X_B,\tau) = 1 - \frac{X_B}{2X_A}
\end{align*}

\subsubsection{Case \#2.3: $X_0 \leq \frac\tau2 \leq X_B \leq \tau$  } 
\begin{align*}
    U_A^*(X_A,X_B,\tau) = & \max_{\substack{ X_1 \geq \tau \\ X_1 \geq X_A} } \crl{ \max_{ \alpha \in \sqr{0,\frac{X_A}{X_1}} } \sqr{ (1-\alpha) \frac{2}{\tau} \prth{ 1 - \frac{X_B}{\tau} } \prth{ \frac{X_A - \alpha X_1}{1-\alpha} } + \alpha \prth{ 1 - \frac{X_B}{X_1 + \sqrt{ X_1^2 - \tau^2 }} } } } \\
    = & \max_{\substack{ X_1 \geq \tau \\ X_1 \geq X_A} } \crl{ \max_{ \alpha \in \sqr{0,\frac{X_A}{X_1}} } \sqr{ \frac{2X_A}{\tau} \prth{ 1 - \frac{X_B}{\tau} } + \alpha \prth{ 1 - \frac{X_B}{X_1 + \sqrt{X_1^2-\tau^2} } - \frac{2X_1}{\tau}\prth{1 - \frac{X_B}{\tau} } } } } \\
    = & \max_{\substack{ X_1 \geq \tau \\ X_1 \geq X_A} } \crl{ \max_{ \alpha \in \sqr{0,\frac{X_A}{X_1}} } \sqr{ \frac{2X_A}{\tau} \prth{ 1 - \frac{X_B}{\tau} } + \alpha \prth{ \frac{X_B\prth{ X_1 + \sqrt{X_1^2 - \tau^2} } + \tau^2 - 2\tau X_1}{\tau^2} } } }
\end{align*}

Note that the constraint $X_0 \leq \frac\tau2$ implies that, 
\begin{align*}
    X_0 = \frac{X_A - \alpha X_1}{1-\alpha} \leq \frac\tau2 \iff \alpha \geq \frac{X_A - \frac\tau2}{X_1 - \frac\tau2}.
\end{align*}

Now, we need to check if the term $X_B\prth{ X_1 + \sqrt{X_1^2 - \tau^2} } + \tau^2 - 2\tau X_1$ is positive or negative. With this in mind, we analyze the equation, 
\begin{align*}
    & X_B\prth{ X_1 + \sqrt{X_1^2 - \tau^2} } + \tau^2 - 2\tau X_1 = 0 \\
    & \sqr{4(X_B-\tau)} X_1^2 + \sqr{ 2\tau( 2\tau - X_B ) } X_1 - \tau\prth{ X_B^2 + \tau^2 } = 0, 
\end{align*}
which has roots, 
\begin{align*}
    \bar{X}_1 = \frac{\tau X_B - 2\tau^2 \pm X_B \sqrt{ \tau(4X_B-3\tau) } }{ 4(X_B - \tau) },
\end{align*}
therefore, the solution of the inner optimization problem depends on if it has or not real roots. 

\subsubsection{Case \#2.3.1: $\frac\tau2 \leq X_B \leq \frac34 \tau$  } 
In this case, the term $X_B\prth{ X_1 + \sqrt{X_1^2 - \tau^2} } + \tau^2 - 2\tau X_1$ is either strictly positive or negative. By checking the vertex of the quadratic equation, 
\begin{align*}
    \sqr{4(X_B-\tau)} X_1^2 + \sqr{ 2\tau( 2\tau - X_B ) } X_1 - \tau\prth{ X_B^2 + \tau^2 } 
\end{align*}
we can check that it is strictly negative. Therefore, we should pick $\alpha$ as small as possible, i.e, 
\begin{align*}
    \alpha^* = \max\crl{ 0, \frac{X_A - \frac\tau2}{X_1 - \frac\tau2} }.
\end{align*}

\subsubsection{Case \#2.3.1.1: $X_A\leq \frac\tau2$ } 
Setting $\alpha^* = 0$ implies that, 
\begin{align*}
    U_A^*(X_A,X_B,\tau) &= \max_{\substack{ X_1 \geq \tau \\ X_1 \geq X_A} } \crl{ \max_{ \alpha \in \sqr{0,\frac{X_A}{X_1}} } \sqr{ \frac{2X_A}{\tau} \prth{ 1 - \frac{X_B}{\tau} } + \alpha \prth{ \frac{X_B\prth{ X_1 + \sqrt{X_1^2 - \tau^2} } + \tau^2 - 2\tau X_1}{\tau^2} } } } \\
    &= \max_{\substack{ X_1 \geq \tau \\ X_1 \geq X_A} } \crl{ \frac{2X_A}{\tau} \prth{ 1 - \frac{X_B}{\tau} } } \\
    &= \frac{2X_A}{\tau} \prth{ 1 - \frac{X_B}{\tau} } 
\end{align*}

\subsubsection{Case \#2.3.1.2: $X_A \geq \frac\tau2$ } 
For this case we have, 
\begin{align*}
    \alpha^* = \frac{X_A - \frac\tau2}{X_1 - \frac\tau2}.
\end{align*}

Then, 
\begin{align*}
    U_A^*(X_A,X_B,\tau) &= \max_{\substack{ X_1 \geq \tau \\ X_1 \geq X_A} } \crl{ \max_{ \alpha \in \sqr{0,\frac{X_A}{X_1}} } \sqr{ \frac{2X_A}{\tau} \prth{ 1 - \frac{X_B}{\tau} } + \alpha \prth{ \frac{X_B\prth{ X_1 + \sqrt{X_1^2 - \tau^2} } + \tau^2 - 2\tau X_1}{\tau^2} } } } \\
    &= \max_{\substack{ X_1 \geq \tau \\ X_1 \geq X_A} } \crl{ \frac{2X_A}{\tau} \prth{ 1 - \frac{X_B}{\tau} } + \prth{ \frac{X_A - \frac\tau2}{X_1 - \frac\tau2} } \prth{ \frac{X_B\prth{ X_1 + \sqrt{X_1^2 - \tau^2} } + \tau^2 - 2\tau X_1}{\tau^2} } }  
\end{align*}

Differentiating with respect to $X_1$,
\begin{align*}
    & \frac{\partial}{\partial X_1} \crl{ \frac{1}{X_1 - \frac\tau2} \prth{ \frac{X_B\prth{ X_1 + \sqrt{X_1^2 - \tau^2} } + \tau^2 - 2\tau X_1}{\tau^2} } } \\
    & = \frac{ X_B \sqr{ \tau\prth{ X_1 + \sqrt{X_1^2-\tau^2} - \frac\tau2 } - \sqrt{X_1^2-\tau^2}\prth{ X_1 + \sqrt{X_1^2-\tau^2} } } }{ \tau \prth{X_1-\frac\tau2}\sqrt{ X_1^2-\tau^2 }\prth{ X_1 + \sqrt{X_1^2-\tau^2} } }
\end{align*}

By first order conditions, 
\begin{align*}
    \tau\prth{ X_1 + \sqrt{X_1^2-\tau^2} - \frac\tau2 } - \sqrt{X_1^2-\tau^2}\prth{ X_1 + \sqrt{X_1^2-\tau^2} } = 0 \implies X_1 = \frac54 \tau.
\end{align*}

Therefore, 
\begin{align*}
    X_1^* = \max\crl{ \frac54 \tau, X_A },
\end{align*}
and,
\begin{align*}
    U_A^*(X_A,X_B,\tau) =  \begin{cases}
        \frac{2X_A}{\tau} \prth{ 1 - \frac{X_B}{\tau} } + \frac{(2X_A-\tau)(4X_B-3\tau)}{3\tau^2}   & \text{ if } X_A \leq \frac54\tau \\
        1 - \frac{X_B}{X_A + \sqrt{X_A^2-\tau^2}}                                                   & \text{ if } X_A \geq \frac54\tau
    \end{cases},
\end{align*}

\subsubsection{Case \#2.3.2: $\frac34\tau \leq X_B \leq \tau$ } 
In this case, the term $X_B\prth{ X_1 + \sqrt{X_1^2 - \tau^2} } + \tau^2 - 2\tau X_1$ could be positive or negative since we have real roots. By checking the vertex of the quadratic equation, 
\begin{align*}
    \sqr{4(X_B-\tau)} X_1^2 + \sqr{ 2\tau( 2\tau - X_B ) } X_1 - \tau\prth{ X_B^2 + \tau^2 } 
\end{align*}
we can check that it is a concave quadratic equation. Therefore, we should pick the biggest $\alpha$ possible in the positive region of the quadratic equation. Therefore, 
\begin{align*}
    \alpha^* = \frac{X_A}{X_1}.
\end{align*}

Then, 
\begin{align*}
    U_A^*(X_A,X_B,\tau) &= \max_{\substack{ X_1 \geq \tau \\ X_1 \geq X_A} } \crl{ \max_{ \alpha \in \sqr{0,\frac{X_A}{X_1}} } \sqr{ \frac{2X_A}{\tau} \prth{ 1 - \frac{X_B}{\tau} } + \alpha \prth{ \frac{X_B\prth{ X_1 + \sqrt{X_1^2 - \tau^2} } + \tau^2 - 2\tau X_1}{\tau^2} } } } \\
    &= \max_{\substack{ X_1 \geq \tau \\ X_1 \geq X_A} } \crl{ \frac{2X_A}{\tau} \prth{ 1 - \frac{X_B}{\tau} } + \prth{ \frac{X_A}{X_1} } \prth{ \frac{X_B\prth{ X_1 + \sqrt{X_1^2 - \tau^2} } + \tau^2 - 2\tau X_1}{\tau^2} } }  \\
    &= \max_{\substack{ X_1 \geq \tau \\ X_1 \geq X_A} } \crl{ \prth{ \frac{X_A}{X_1} } \prth{ 1 -\frac{X_B}{X_1+ \sqrt{X_1^2-\tau^2}} } } .
\end{align*}

Similar to the case when $\tau \leq X_B$, we have that, 
\begin{align*}
    X_1^* = \max\crl{ \sqrt{X_B^2+\tau^2}, X_A },
\end{align*}
and,
\begin{align*}
    U_A^*(X_A,X_B,\tau) =  \begin{cases}
        \frac{X_A}{X_B + \sqrt{ X_B^2 + \tau^2 } }             & \text{ if } \sqrt{X_B^2+\tau^2} \geq X_A \\
        1 - \frac{X_B}{X_A + \sqrt{X_A^2-\tau^2}}              & \text{ if } \sqrt{X_B^2+\tau^2} \leq X_A
    \end{cases}
\end{align*}

\subsubsection{Case \#2.4: $X_0 \leq X_B \leq \frac\tau2$ } 
\begin{align*}
    U_A^*(X_A,X_B,\tau) = & \max_{\substack{ X_1 \geq \tau \\ X_1 \geq X_A} } \crl{ \max_{ \alpha \in \sqr{0,\frac{X_A}{X_1}} } \sqr{ (1-\alpha) \frac{1}{2X_B} \prth{ \frac{X_A - \alpha X_1}{1-\alpha} } + \alpha \prth{ 1 - \frac{X_B}{X_1 + \sqrt{ X_1^2 - \tau^2 }} } } } \\
    = & \max_{\substack{ X_1 \geq \tau \\ X_1 \geq X_A} } \crl{ \max_{ \alpha \in \sqr{0,\frac{X_A}{X_1}} } \sqr{ \frac{X_A}{2X_B} + \alpha \prth{ 1 - \frac{X_B}{X_1 + \sqrt{ X_1^2 - \tau^2 }}  - \frac{X_1}{2X_B} } } } 
\end{align*}

The constraint $X_0 \leq X_B$ implies that,
\begin{align*}
    X_0 = \frac{X_A - \alpha X_1}{1-\alpha} \leq X_B \iff \frac{X_A-X_B}{X_1-X_B} \leq \alpha.
\end{align*}

Now, we need to know if the term, 
\begin{align} \label{eq:q2.4}
    1 - \frac{X_B}{X_1 + \sqrt{ X_1^2 - \tau^2 }}  - \frac{X_1}{2X_B} 
\end{align}
is positive or negative. With that in mind, we analyze when it is equal to $0$, 
\begin{align*}
    1 - \frac{X_B}{X_1 + \sqrt{ X_1^2 - \tau^2 }}  - \frac{X_1}{2X_B}  = 0 \iff \sqr{ - 4X_B^2-\tau^2 } X_1^2 + \sqr{ 4X_B\prth{\tau^2+2X_B^2} } X_1 + \sqr{ (-4X_B^2)\prth{ X_B^2+\tau^2 } } = 0.
\end{align*}

We can check that the discriminant of the previous quadratic equation $-\tau^2X_B^2$ is strictly negative. Then, it does not have real solution. Then, Equation~\eqref{eq:q2.4} can only be strictly positive or strictly negative. By checking its vertex, we can ensure that, 
\begin{align*} 
    1 - \frac{X_B}{X_1 + \sqrt{ X_1^2 - \tau^2 }}  - \frac{X_1}{2X_B} \leq 0.
\end{align*}
Then, we need to pick the smallest $\alpha$ possible, 
\begin{align*}
    \alpha^* = \max\crl{ 0, \frac{X_A-X_B}{X_1-X_B} }
\end{align*}

\subsubsection{Case \#2.4.1: $X_A\leq X_B$ } 
Setting $\alpha^* = 0$ implies that, 
\begin{align*}
    U_A^*(X_A,X_B,\tau) &= \max_{\substack{ X_1 \geq \tau \\ X_1 \geq X_A} } \crl{ \max_{ \alpha \in \sqr{0,\frac{X_A}{X_1}} } \sqr{ \frac{X_A}{2X_B} + \alpha \prth{ 1 - \frac{X_B}{X_1 + \sqrt{ X_1^2 - \tau^2 }}  - \frac{X_1}{2X_B} } } } \\
    &= \max_{\substack{ X_1 \geq \tau \\ X_1 \geq X_A} } \crl{ \frac{X_A}{2X_B}  } \\
    &= \frac{X_A}{2X_B} 
\end{align*}

\subsubsection{Case \#2.4.2: $X_B \leq X_A$ } 
The solution for this case have been already analyzed in the previous cases; particularly, in case 2.2.

\subsubsection{ Definition of $U_A^*(X_A,X_B,\tau)$ }
We can use the solutions for all the cases to recover a definition for $U_A^*(X_A,X_B,\tau)$ for the entire $(X_A,X_B,\tau)$ space as follows, 
\begin{itemize}
    \item $\tau \leq X_B$. Case \#1.
    \item $\frac34\tau \leq X_B \leq \tau$. Case \#2.3.2.
    \item $\frac\tau2 \leq \frac34\tau$. Case \#2.3.1. 
    \begin{itemize}
        \item $X_A \leq \frac\tau2$. Case \#2.3.1.1.
        \item $X_A \geq \frac\tau2$. Case \#2.3.1.2.
    \end{itemize}
    \item $X_B \leq \frac\tau2$ 
    \begin{itemize}
        \item $X_A \leq X_B \leq \frac\tau2$. Case \#2.4.1.
        \item $X_B \leq X_A \leq \frac\tau2$. Case \#2.4.2.
        \item $X_B \leq \frac\tau2 \leq X_A$. Cases \#2.1 and \#2.2.2.
    \end{itemize}
\end{itemize}
With the expression for each cases we obtain the definition for $U_A^*(X_A,X_B,\tau)$. To obtain the definition in Equations~\eqref{eq:ua:i}-\eqref{eq:ua:ix} we need to use the fact that $U_B^*(X_A,X_B,\tau) = 1-U_A^*(X_A,X_B,\tau)$.

%%%%%%%%%%%%%%%%%%%%%%%%%%%%%%%%%%%%%%%%%%%%%%%%%%%%%%%%%%%%%%%%%%%%%%%%%%%%%%%%%%%%%%%%%%%%%%%%%%%%%%%%%%%%%%%%%%%%%%%%%%%%%%%%%%%%%%%%%%%%%%%%%%%%%%%%%%%%%%%%%%%%%%%%%%%%%%%%%%%%%%%%%%%%%%%%%%%%%%%%%%%%%%%%%%%%%%%%%%%%%%%%%%%%%%%%%%%%%%%%%%%%%%%%%%%%%%%%%

\end{document}

%% file: def.tex
\usepackage{ntheorem}
\usepackage{hyperref}
\usepackage{textcomp}
\let\labelindent\relax
\usepackage{enumitem}
\usepackage{graphicx}
\usepackage{epstopdf}
\usepackage{float}
\usepackage[nocomma]{optidef}
\usepackage{dsfont}
\usepackage{times}
\usepackage{bm}
\usepackage{fancyhdr,graphicx,amsmath,amssymb}
\usepackage[caption=false,font=footnotesize]{subfig}
\usepackage{cite}
\usepackage{color}
\usepackage[algo2e,ruled,vlined]{algorithm2e}
\usepackage{placeins}
\usepackage{cases}

\usepackage{multirow}
\usepackage{makecell}

\newtheorem{theorem}{Theorem}

\newtheorem{corollary}{Corollary}
\newtheorem{lemma}{Lemma}

\newtheorem{definition}{Definition}

\newcommand{\prth}[1]{\left( #1 \right) }
\newcommand{\sqr}[1]{\left[ #1 \right] }
\newcommand{\crl}[1]{\left\{ #1 \right\} }

\newcommand{\F}{\mathbb{F}}
\newcommand{\G}{\mathcal{G}}
\newcommand{\I}{\mathbb{I}}

\newcommand{\R}{\mathbb{R}}

\renewcommand{\Pr}{\mathbb{P}}
\newcommand{\Ex}{\mathbb{E}}

\newcommand{\Ind}{\mathbf{1}}

\newcommand{\eps}{\varepsilon}

\DeclareMathOperator*{\argmin}{arg\,min}

\DeclareMathOperator{\supp}{supp}

\DeclareMathOperator{\GL}{GL}
\DeclareMathOperator{\GLI}{GLI}